# Hierarchical Sparse Modeling: A Choice of Two Group Lasso Formulations

**Xiaohan Yan and Jacob Bien**

*Abstract.* Demanding sparsity in estimated models has become a routine practice in statistics. In many situations, we wish to require that the sparsity patterns attained honor certain problem-specific constraints. *Hierarchical sparse modeling* (HSM) refers to situations in which these constraints specify that one set of parameters be set to zero whenever another is set to zero. In recent years, numerous papers have developed convex regularizers for this form of sparsity structure, which arises in many areas of statistics including interaction modeling, time series analysis, and covariance estimation. In this paper, we observe that these methods fall into two frameworks, the *group lasso* (GL) and *latent overlapping group lasso* (LOG), which have not been systematically compared in the context of HSM. The purpose of this paper is to provide a side-by-side comparison of these two frameworks for HSM in terms of their statistical properties and computational efficiency. We call special attention to GL's more aggressive shrinkage of parameters deep in the hierarchy, a property not shared by LOG. In terms of computation, we introduce a finite-step algorithm that exactly solves the proximal operator of LOG for a certain simple HSM structure; we later exploit this to develop a novel path-based block coordinate descent scheme for general HSM structures. Both algorithms greatly improve the computational performance of LOG. Finally, we compare the two methods in the context of covariance estimation, where we introduce a new sparsely-banded estimator using LOG, which we show achieves the statistical advantages of an existing GL-based method but is simpler to express and more efficient to compute.

## 1. INTRODUCTION

Convex regularizers for sparse modeling are ubiquitous in the statistics and machine learning literatures. Regularizers such as the *lasso* (Tibshirani, 1996) and the *group lasso* (Turlach, Venables and Wright, 2005, Yuan and Lin, 2006) are commonly-used tools for seemlessly integrating model selection into statistical procedures, thereby extending these methods' reach to high-dimensional settings in which the number of parameters greatly exceeds the sample size. In contrast to the lasso, which seeks sparsity with no *a priori* pattern, the group lasso regularizer allows pre-defined groups of variables to be set to zero simultaneously, giving rise to the so-called *structured sparsity* literature in which certain patterns of zeros are sought (Bach et al., 2012). The focus of this paper is on a particular kind of structured sparsity that arises in many statistics problems, which we will call *hierarchical sparse modeling* (HSM). Given a vector $\beta \in \mathbb{R}^p$ of parameters and a known collection of non-empty, disjoint sets $s_1, \dots, s_N \subseteq \{1, \dots, p\}$, HSM focuses on situations in which we wish to set groups of variables to zero while

*Xiaohan Yan is Ph.D. Candidate, Department of Statistical Science, Cornell University, Comstock Hall, Ithaca, New York 14853, USA (e-mail: xy257@cornell.edu). Jacob Bien is Assistant Professor, Department of Data Sciences and Operations, University of Southern California, Bridge Hall, Los Angeles, California 90089, USA (e-mail: jbien@usc.edu).*

ensuring that

$$\beta_{s_i} = 0 \implies \beta_{s_j} = 0$$

for certain ordered pairs of groups $(s_i, s_j)$. More specifically, in HSM one forms a directed acyclic graph (DAG) over $\{s_1, \dots, s_N\}$ to encode the desired hierarchical sparsity relations (one requires the above to hold if $s_i$ is an ancestor of $s_j$ in the DAG). HSM appears in many applications in statistics, including interactions (Yuan, Joseph and Zou, 2009, Zhao, Rocha and Yu, 2009, Radchenko and James, 2010, Schmidt and Murphy, 2010, Choi, Li and Zhu, 2010, Jenatton et al., 2010, Bien, Taylor and Tibshirani, 2013, Lim and Hastie, 2015, She, Wang and Jiang, 2017, Haris, Witten and Simon, 2016), covariance matrix estimation (Levina, Rothman and Zhu, 2008, Rothman, Levina and Zhu, 2010, Bien, Bunea and Xiao, 2016), additive models (Lou et al., 2016, Chouldechova and Hastie, 2015), time series models (Nicholson, Bien and Matteson, 2014), and multiple kernel learning (Bach, 2008). We note that *hierarchical sparse coding* is a common special case of HSM in which the DAG is a forest of trees (Zhao, Rocha and Yu, 2009, Jenatton et al., 2011). For example, in a two-way interaction model of the form

$$Y = \beta_0 + \beta_1 X_1 + \beta_2 X_2 + \beta_3 X_3 + \beta_{12} X_1 X_2 + \beta_{13} X_1 X_3 + \beta_{23} X_2 X_3 + \varepsilon,$$

one can express the principle of marginality (Nelder, 1977) as that $\beta_j$ and $\beta_k$ are parents of $\beta_{jk}$ (each node of the DAG contains a single element, that is, $|s_i| = 1$ for all $i$). The DAG, which is not a tree, is depicted in Figure 1. A simpler DAG structure arises in banded covariance estimation, in which a $p \times p$ matrix $\Sigma$'s sparsity pattern can be described by having the elements of each subdiagonal set to zero only if those farther from the main diagonal than it are also all set to zero (in this situation, the DAG is simply a path as depicted in Figure 4 with $D = p - 1$). We will discuss banded covariance estimation in greater detail in Section 5.

There are two primary convex regularizers used for structured sparsity: the *group lasso* (GL) and *latent overlapping group lasso* (LOG) (Jacob, Obozinski and Vert, 2009). The sparsity patterns attained by these regularizers are in general different in nature, and so the regularizers typically arise in complementary situations. Given a set of groups of parameters $\mathcal{G}$, GL *sets to zero a union of groups* that is a subset of $\mathcal{G}$. The GL penalty is defined as a weighted sum of $\ell_2$ norms over groups of parameters as defined in $\mathcal{G}$:

$$\Omega^{\mathcal{G}}_{\mathrm{GL}}(\beta; w) = \sum_{g \in \mathcal{G}} w_g \|\beta_g\|_2. \tag{1}$$

Here, $w_g$ are positive scalars that control the relative strength of the terms within the GL penalty.

Jacob, Obozinski and Vert (2009) observe that when the groups in $\mathcal{G}$ overlap, the induced support from GL may not be a union of groups since the complement of a union of groups is not necessarily a union of groups. In this sense, the group lasso as defined in (1) should not be used in situations in which one wishes a subset of (overlapping) groups to remain nonzero. The authors propose LOG as a solution to this problem. Rather than apply the $\ell_1/\ell_2$ norm directly on the parameter vector $\beta$, LOG forms the parameters as a sum of GL-penalized latent variables, which is each supported by a group $g$:

$$\Omega^{\mathcal{G}}_{\mathrm{LOG}}(\beta; w) = \inf_{\{v^{(g)} \in \mathbb{R}^p\}_{g \in \mathcal{G}}} \Big\{ \sum_{g \in \mathcal{G}} w_g \|v^{(g)}\|_2 \text{ s.t. } \sum_{g \in \mathcal{G}} v^{(g)} = \beta \text{ and } v^{(g)}_{g^c} = 0 \text{ for } g \in \mathcal{G} \Big\}. \tag{2}$$

In LOG, a subset of the latent variables is set to zero. Since $\beta$ is formed as a sum of these latent variables, the parameters in a group $g$ are selected as long as the corresponding latent variable $v^{(g)}$ is nonzero. As a result, the LOG penalty *leaves nonzero a union of groups*.

Although GL and LOG induce different sparsity patterns in general, we show in Section 2 that in the special case of HSM, either regularizer (with an appropriately chosen group structure) can be used to accomplish the HSM structure. From a methodological statistician's standpoint, this observation leads to ambiguity as to which regularizer one should use for HSM. Indeed, a survey of the HSM literature reveals that researchers have been using both frameworks with no discussion of the seemingly arbitrary choice about whether to use GL or LOG. Table 1 arranges methods developed across five statistical domains according to which regularizer was used. One observes that LOG is the less commonly employed regularizer in HSM problems. The objective of this paper is to compare the GL and LOG approaches in the context of HSM. While the class of sparsity patterns obtainable is the same for the two regularizers, we show in Section 2.3 that the nature of the shrinkage is different even for the simplest nontrivial HSM problem.

The main contributions of our investigation into these two regularizers are summarized below:

- In Section 3, we show that the GL penalty as defined in (1) tends to apply a greater amount of shrinkage to parameters embedded deep in the DAG whereas LOG does not. In certain situations where this more aggressive shrinkage is not desired, a more complicated weighting scheme can be adopted (as was done in Jenatton, Audibert and Bach, 2011, Bien, Bunea and Xiao, 2016). This weighting scheme, which makes computation and theory more involved, appears to be necessary to match the statistical performance of LOG.

TABLE 1
*Applications of GL and LOG in HSM*

| **Problem** | **Group lasso (GL)** | **Latent overlapping GL (LOG)** |
|---|---|---|
| Hierarchical interactions | CAP, Zhao, Rocha and Yu (2009)<br>VANISH, Radchenko and James (2010),<br>Schmidt and Murphy (2010)<br>hiernet, Bien, Taylor and Tibshirani (2013)<br>GRESH, She, Wang and Jiang (2017)<br>FAMILY, Haris, Witten and Simon (2016) | glinternet, Lim and Hastie (2015) |
| Banded covariance matrix | hierband, Bien, Bunea and Xiao (2016) | Section 5 of this paper |
| Generalized partially linear additive models | SPLAM, Lou et al. (2016) | GAMSel, Chouldechova and Hastie (2015) |
| Times series | HVAR, Nicholson, Bien and Matteson (2014) | — |
| Hierarchical multiple kernel learning | HKL, Bach (2008) | — |

- In Section 4, we focus on computational aspects. It was shown in Jenatton et al. (2011) that when the DAG is a tree, the proximal operator of GL could be solved exactly in a finite number of operations. While there is no known corresponding algorithm for LOG, in the special case that the DAG is a path graph (or forest of path graphs), we derive such an algorithm. We then leverage this result to introduce a novel path-based block coordinate descent (BCD) scheme for the case of a general DAG that is more efficient than the standard BCD algorithm.
- In Section 5, as a case study, we demonstrate how the LOG framework can be used instead of GL for the problem of estimating a banded covariance matrix. We use banded covariance matrix estimation as a primary basis to compare the statistical performance between the GL and LOG frameworks. We prove that this estimator attains the same bandwidth recovery properties and convergence rate as the "convex banding" estimator of Bien, Bunea and Xiao (2016), which had to rely on a complicated weighting scheme. Furthermore, we find that it attains similar empirical performance.

### 1.1 Notation

We use $\|\beta\|_2$ and $\|\mathbf{\Sigma}\|_F$ for the $\ell_2$ norm of a vector $\beta \in \mathbb{R}^p$ and the Frobenius norm of a matrix $\mathbf{\Sigma} \in \mathbb{R}^{p\times p}$, respectively. The support of $\beta$ is denoted $\operatorname{supp}(\beta) \subseteq \{1,\dots,p\}$, which is the set of indices of nonzero elements in $\beta$. For $\beta$, a group of parameters is a subset $g \subseteq \{1,\dots,p\}$. We use $\mathcal{G}$ to denote the set of groups. The weight vector $w$, of the same size as $\mathcal{G}$, has positive elements. For a group $g \subseteq \{1,\dots,p\}$, $\beta_g \in \mathbb{R}^p$ has the same entries as $\beta$ for indices in $g$ and is 0 for all other indices, whereas $\beta_{|g} \in \mathbb{R}^{|g|}$ is a subset of $\beta$ for indices in $g$. For a matrix $\mathbf{X} \in \mathbb{R}^{n\times p}$ and a subset $g \subseteq \{1,\dots,p\}$, $\mathbf{X}_{|g} \in \mathbb{R}^{n\times|g|}$ has the same columns as $\mathbf{X}$ for column indices in $g$. In Section 5, given a subset of a matrix indices $g \subseteq \{1,\dots,p\}^2$ of a matrix $\mathbf{\Sigma}$, let $\mathbf{\Sigma}_g \in \mathbb{R}^{p\times p}$ be a matrix whose entries are the same as $\mathbf{\Sigma}$ for the indices in $g$, and are 0 for other indices. Let $(\cdot)_+ = \max\{\cdot, 0\}$ denote the positive part and $S(\cdot,\cdot)$ and $S_G(\cdot,\cdot)$ the elementwise and groupwise soft-thresholding operators, respectively,

$$[S(y,\mu)]_i = y_i\left(1-\frac{\mu}{|y_i|}\right)_+ \quad \text{and}$$

$$S_G(y,\mu) = y\left(1-\frac{\mu}{\|y\|}\right)_+,$$

where $\|\cdot\|$ denotes $\|\cdot\|_2$ or $\|\cdot\|_F$, depending on whether $y$ is a vector or a matrix.

## 2. HIERARCHICAL SPARSE MODELING: TWO FRAMEWORKS

Let $s_1,\dots,s_N \subseteq \{1,\dots,p\}$ be a collection of nonempty, disjoint sets of indices and let $\mathcal{D}$ be a DAG with vertex set $\{s_1,\dots,s_N\}$. In specifying a DAG, the notions of *ancestor* and *descendant* are well defined. In particular, we let $\operatorname{descendants}(\mathcal{D}; s_i)$ denote the set of all $s_j$ for which there exists a path from $s_i$ to $s_j$ in $\mathcal{D}$ and we likewise let ancestors$(\mathcal{D}; s_j)$ denote the set of all $s_i$ for which there exists a path from $s_i$ to $s_j$. *Note that we let a node itself be in both its ancestor group and its descendant group.* To better illustrate the constructions of *ancestor* and *descendant*, we use a two-way interaction model with three predictors as an example. The corresponding DAG for the interaction model is shown in Figure 1. To be specific, for each main effect $\beta_j$, the two interaction effects resulted from $\beta_j$ and another main effect $\beta_k$ are considered as descendancts of $\beta_j$. Conversely, for the interaction effect $\beta_{jk}$, its two parent main effects, $\beta_j$ and $\beta_k$, are its ancestors.

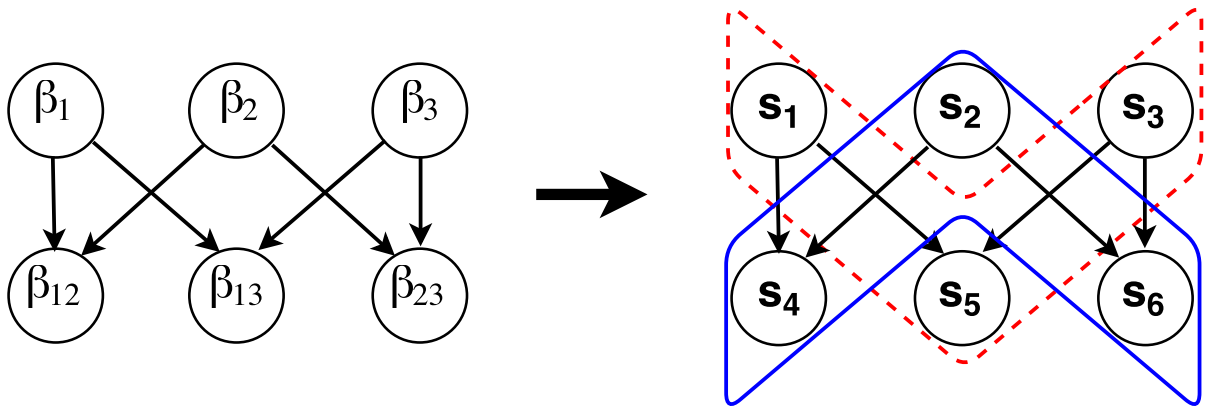

FIG. 1. *Left*: *A DAG $\mathcal{D}$ for a two-way interaction model with three predictors. In HSM*, *the DAG $\mathcal{D}$ encodes the sparsity structure*: *a node's parameters must be set to zero if it has a parent with zeroed parameters. Right*: *The same $\mathcal{D}$ specified using our notation*: *each node contains only one element and the correspondence between $s_i$ and $\beta_j$ is as shown. In red dashed contour*, ancestors$(\mathcal{D}; s_5) = \{s_1, s_3, s_5\}$ *include both main effects*, $\beta_1$ *and* $\beta_3$, *in the ancestor group of the interaction effect* $\beta_{13}$. *In blue solid contour*, descendants$(\mathcal{D}; s_2) = \{s_2, s_4, s_6\}$ *contains both interaction effects involving main effect* $\beta_2$.

The goal of HSM is to attain sparsity patterns for which

$$\beta_{s_i} = 0 \quad \Rightarrow \quad \beta_{s_j} = 0 \quad \text{for all } s_j \in \text{descendants}(\mathcal{D}; s_i). \tag{3}$$

In the context of our interaction model example, (3) enforces the selection that all the resulting interaction effects are discarded if the main effect is not selected. We can equivalently express (3) as

$$\beta_{s_j} \neq 0 \quad \Rightarrow \quad \beta_{s_i} \neq 0 \quad \text{for all } s_i \in \text{ancestors}(\mathcal{D}; s_j). \tag{4}$$

In interaction modeling, this tells us that all its parent main effects need to be selected if an interaction effect is selected. Given (3) and (4) are functionally equivalent statements, we show in Sections 2.1 and 2.2 how GL and LOG are based on (3) and (4), respectively. While their sparsity patterns are equivalent, we show in Section 2.3 that the two approaches lead to different solutions.

### 2.1 The Group Lasso Approach

To induce the hierarchical sparsity of (3), Zhao, Rocha and Yu (2009), Jenatton et al. (2011) and many others use the GL regularizer (1) with group structure $\mathcal{G}$ chosen to be

$$d(\mathcal{D}) := \{\text{descendants}(\mathcal{D}; s_i) : i = 1, \dots, N\}. \tag{5}$$

The top panels of Figure 2 gives an example of $d(\mathcal{D})$ for a DAG associated with a two-way interaction model with three predictors. There is a group corresponding to each node $s_i$, and this group contains all the parameters in $s_i$ and in its descendant nodes. Recalling that GL sets to zero a union of groups, we see that $\Omega_{\mathrm{GL}}^{d(\mathcal{D})}$ achieves (3). As shown in the top panels of Figure 2, each main effect is grouped with its descendant interaction effects, whereas each interaction effect is grouped by itself. It is possible for an interaction effect to be zeroed out while keeping its parent main effects significant. However, whenever the main effect is zeroed out which only occurs when the whole group (including interaction effects) is not selected, all the descendant interaction effects must be zeroed out as well. We choose a convex smooth loss function $F$ depending on the statistical context (a common choice is the negative log-likelihood) and then solve

$$\min_{\beta \in \mathbb{R}^p} \{F(\beta) + \lambda \Omega_{\mathrm{GL}}^{d(\mathcal{D})}(\beta; w)\}. \tag{6}$$

Here, $\lambda \geq 0$ is a regularization parameter that controls the sparsity level of $\beta$.

### 2.2 The Latent Overlapping Group Lasso Approach

The LOG penalty (2) of Jacob, Obozinski and Vert (2009) can be used for HSM taking the perspective of (4). We choose $\mathcal{G}$ to be

$$a(\mathcal{D}) := \{\text{ancestors}(\mathcal{D}; s_j) : j = 1, \dots, N\}. \tag{7}$$

For each node $s_j$ in $\mathcal{D}$, there is a group containing all parameters that are contained in $s_j$ or its ancestors. The bottom panels of Figure 2 shows $a(\mathcal{D})$ for the same DAG as on the top. As observed in Jacob, Obozinski and Vert (2009), LOG leaves a union of groups nonzero, thus we see that (4) is accomplished by $\Omega_{\mathrm{LOG}}^{a(\mathcal{D})}$. In our interaction model example, as shown in the bottom panels of Figure 2, each interaction effect is grouped with both parent main effects, whereas each main effect is grouped by itself separately. This group structure guarantees (4) since both main effects will be recovered as nonzero if we have a nonzero interaction effect, given they are in the same group. We are thus faced with a choice of whether to use an estimator defined based on solving (6) versus one based on solving

$$\min_{\beta \in \mathbb{R}^p} \{F(\beta) + \lambda \Omega_{\mathrm{LOG}}^{a(\mathcal{D})}(\beta; w)\}. \tag{8}$$

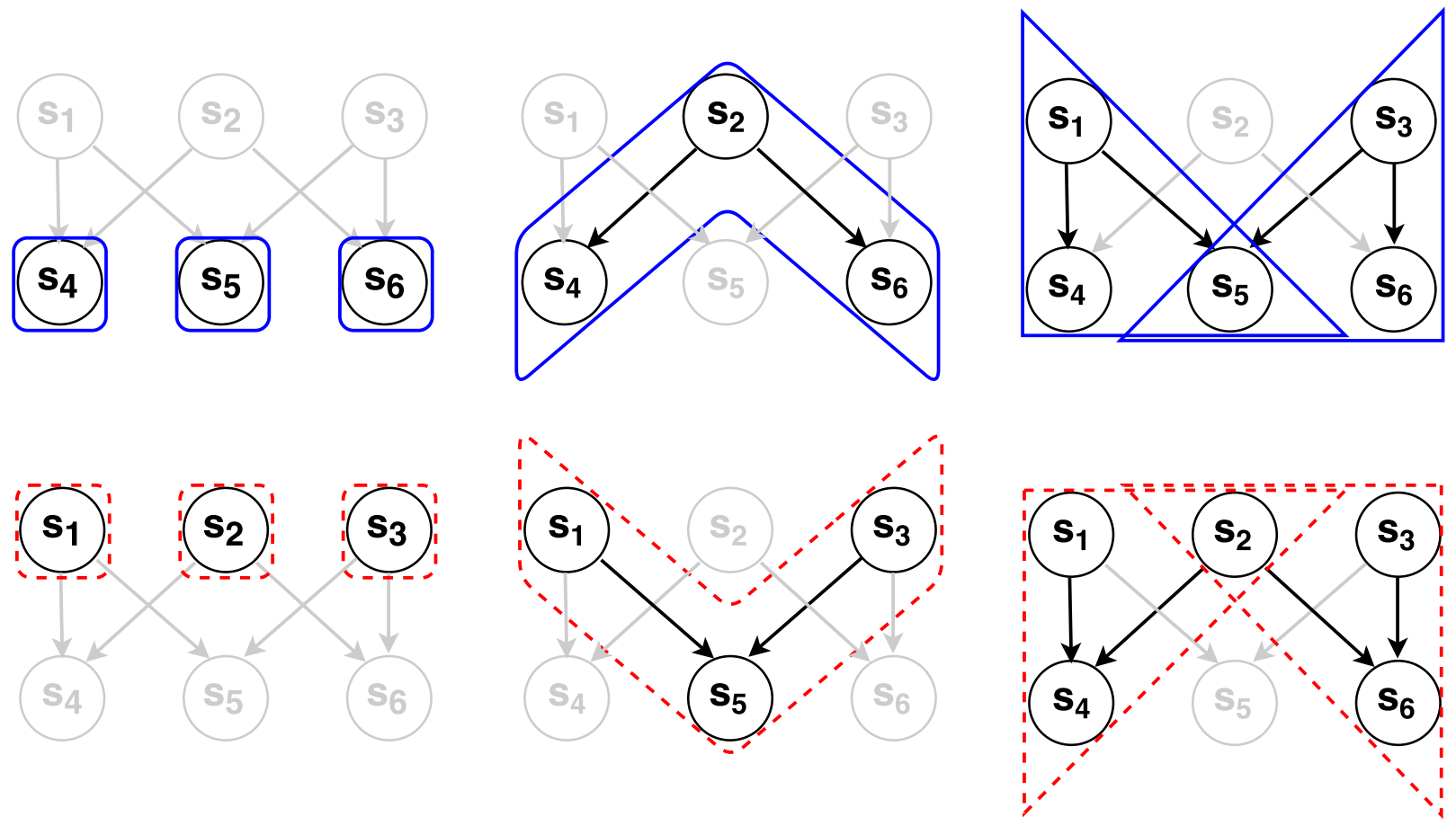

FIG. 2. *For the same DAG as in Figure 1, an illustration of group structures $\mathcal{G} = d(\mathcal{D})$ and $\mathcal{G} = a(\mathcal{D})$ induced for GL and LOG, respectively. Top: The group structure $d(\mathcal{D})$ for GL is shown in solid contours: $d(\mathcal{D}) = \{s_4, s_5, s_6, s_2 \cup s_4 \cup s_6, s_1 \cup s_4 \cup s_5, s_3 \cup s_5 \cup s_6\}$. Each group of $d(\mathcal{D})$ can be thought of as a set of the effect itself and all the relevant interaction effects. Bottom: The group structure $a(\mathcal{D})$ for LOG is shown in dashed contours: $a(\mathcal{D}) = \{s_1, s_2, s_3, s_1 \cup s_3 \cup s_5, s_1 \cup s_2 \cup s_4, s_2 \cup s_3 \cup s_6\}$. Each group of $a(\mathcal{D})$ can be described as a set of the effect itself and all the relevant main effects.*

### 2.3 Are These Two Approaches Different?

In Sections 2.1 and 2.2, we describe two frameworks that lead to the same set of sparsity patterns. This equivalence can be shown geometrically in the simple case in which $p = 3$, $s_i = \{i\}$ for $i = 1, 2, 3$, and $\mathcal{D}$ is the path graph $s_1 \to s_2 \to s_3$. Figure 3 depicts the unit ball of the induced GL and LOG penalties introduced in the previous sections. We observe that both balls have their nondifferentiable points lying in the plane defined by $\beta_3 = 0$. Furthermore, both unit balls have "poles" on the axis defined by $\beta_2 = \beta_3 = 0$. Given that both penalties lead to the same set of supports, it is natural to ask if these two regularizers are in fact identical for an appropriately chosen set of weights. We consider the simplest nontrivial HSM: let $p = 2$, $s_1 = \{1\}$ and $s_2 = \{2\}$, and take $\mathcal{D}$ to be a single edge connecting singleton sets: $s_1 \to s_2$. The following lemma establishes that these two penalties are different even in this simplest of situations.

LEMMA 1. *Take $\mathcal{D}$ to be $\{1\} \to \{2\}$ and fix $w' = (1, 1)$. There does not exist $w \in \mathbb{R}^{+2}$ such that*

$$\Omega_{\mathrm{GL}}^{d(\mathcal{D})}(\beta; w) = \Omega_{\mathrm{LOG}}^{a(\mathcal{D})}(\beta; w') \quad \forall \beta \in \mathbb{R}^2.$$

PROOF. See Appendix A. □

Moreover, we can compare the proximal operators of the two penalties, which correspond to (6) and (8) with $F(\beta) = \frac{1}{2}\|y - \beta\|_2^2$:

$$\begin{aligned} &\mathrm{Prox}_{\mathrm{GL}}^{d(\mathcal{D})}(y; \lambda, w) \\ &\quad := \arg\min_{\beta \in \mathbb{R}^p} \left\{ \frac{1}{2}\|y - \beta\|_2^2 + \lambda \Omega_{\mathrm{GL}}^{d(\mathcal{D})}(\beta; w) \right\}, \end{aligned} \tag{9}$$

$$\begin{aligned} &\mathrm{Prox}_{\mathrm{LOG}}^{a(\mathcal{D})}(y; \lambda, w) \\ &\quad := \arg\min_{\beta \in \mathbb{R}^p} \left\{ \frac{1}{2}\|y - \beta\|_2^2 + \lambda \Omega_{\mathrm{LOG}}^{a(\mathcal{D})}(\beta; w) \right\}. \end{aligned} \tag{10}$$

The use of equality in the above definition is justified by observing that $F$ is strongly convex and therefore the arg min is a single point. The path graph structure of the simplest HSM example allows us to express both proximal operators in closed form, which allows us to see plainly how they differ. Let $\hat{\beta}^{\mathrm{GL}}$ and $\hat{\beta}^{\mathrm{LOG}}$ denote the solution to the respective proximal operators defined in (9) and (10).

LEMMA 2. *Taking $\mathcal{D}$ to be $\{1\} \to \{2\}$, $\hat{\beta}^{\mathrm{GL}}$ and $\hat{\beta}^{\mathrm{LOG}}$ can be written in closed form:*

$$\hat{\beta}^{\mathrm{GL}} = S_G\left(\begin{pmatrix} y_1 \\ S(y_2, \lambda w_2) \end{pmatrix}, \lambda w_1\right),$$

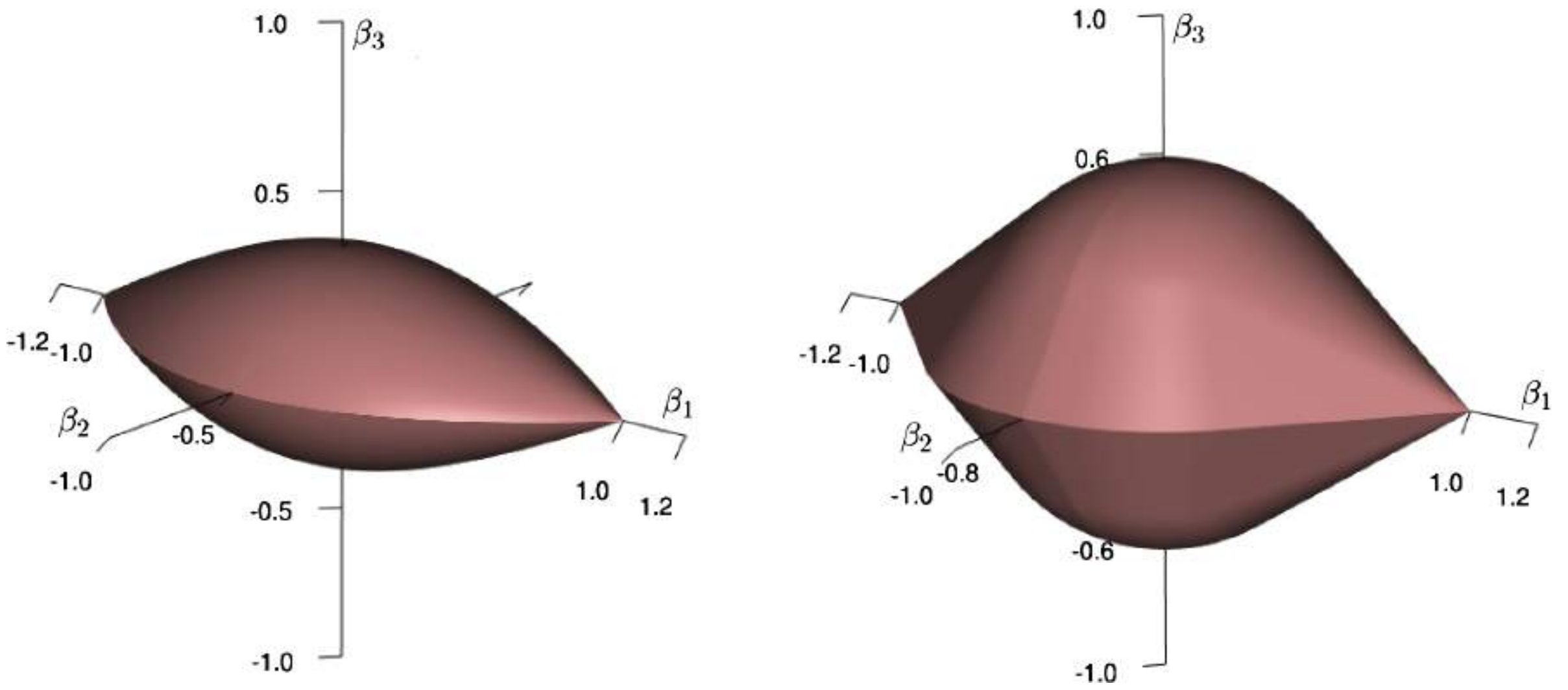

FIG. 3. *For $\beta \in \mathbb{R}^3$ and the DAG $\{1\} \to \{2\} \to \{3\}$, left: the unit ball of $\Omega_{\mathrm{GL}}^{d(\mathcal{D})}(\beta; w)$ where $d(\mathcal{D}) = \{\{1,2,3\},\{2,3\},\{3\}\}$ and $w = (1,1,1)$, and right: the unit ball of $\Omega_{\mathrm{LOG}}^{a(\mathcal{D})}(\beta; w)$ where $a(\mathcal{D}) = \{\{1\},\{1,2\},\{1,2,3\}\}$ and $w = (1, \sqrt{2}, \sqrt{3})$.*

$$\hat{\beta}^{\mathrm{LOG}} = \begin{cases} S_G(y, \lambda w_2) \\ \quad \text{if } |y_2| \geq \dfrac{\sqrt{w_2^2 - w_1^2}}{w_1}|y_1|, \\ \begin{pmatrix} S(y_1, \lambda w_1) \\ S\Big(y_2, \lambda\sqrt{w_2^2 - w_1^2}\Big) \end{pmatrix} \\ \quad \text{otherwise} \end{cases}$$

*with $w_1$ and $w_2$ in GL being applied on the group $\{1, 2\}$ and $\{2\}$, respectively, and $w_1$ and $w_2$ in LOG being applied on the group $\{1\}$ and $\{1, 2\}$, respectively.*

PROOF. This result follows by applying Algorithms 1 and 3 in Section 4. □

We see that $\hat{\beta}_2^{\mathrm{GL}}$ has two "chances" to be set to zero: first, through the elementwise soft thresholding of $y_2$ and, second, through the groupwise soft-thresholding of $(y_1, S(y_2, \lambda w_2))$. By contrast, for $\hat{\beta}_2^{\mathrm{LOG}}$, the shrinkage is applied only once (though whether it is an elementwise or groupwise soft-thresholding depends on the relative size of $|y_1|$ and $|y_2|$). This example establishes that these two regularizers are in fact different, so we proceed to investigate the nature and implications of this difference.

## 3. DIFFERENTIAL SHRINKAGE OF GL

In this section, we call attention to a property of the GL shrinkage that is not shared by LOG: namely, that $\Omega_{\mathrm{GL}}^{d(\mathcal{D})}$ shrinks parameters embedded in nodes deep in the DAG $\mathcal{D}$ more agressively than those that are in less deep nodes in the DAG. This "over-penalization" phenomenon has been observed previously (Jenatton, Audibert and Bach, 2011, Bach et al., 2012, Bien, Bunea and Xiao, 2016) in overlapping group lasso settings, but it does not appear to be widely appreciated. A simple explanation for this phenomenon is that the vector $\beta_{s_j}$ appears within $\Omega_{\mathrm{GL}}^{d(\mathcal{D})}$ in $|\text{ancestors}(\mathcal{D}; s_j)|$ terms, a number that can vary greatly among different $s_j$. In Section 4, we will see that the amount of shrinkage of $\beta_{s_j}$ grows with the number of groups its indices $s_j$ belong to. For example, for the path graph $\mathcal{D}$ shown in Figure 4, $\beta_{s_1}$ appears in only a single groupwise soft-thresholding whereas $\beta_{s_D}$ is soft-thresholded $D$ times. The uneven distribution of shrinkage over the support in GL is a nonnegligible phenomenon. By contrast, we will show that $\Omega_{\mathrm{LOG}}^{a(\mathcal{D})}$ applies a comparable amount of shrinkage at all depths of $\mathcal{D}$.

In order to more directly study the difference of the shrinking mechanisms in GL and LOG, we will compare the solutions to (9) and (10) for the directed path graph in Figure 4 in the case that there is one parameter per node, that is, $s_i = \{i\}$ for $i = 1, \ldots, D$. For simplicity, we consider $y \sim N_D(\beta^*, \sigma^2 I_D)$ where $\beta^*$ is an unknown mean vector. The group structure $d(\mathcal{D})$ for GL for this DAG consists of groups of the form $\{i, \ldots, D\}$

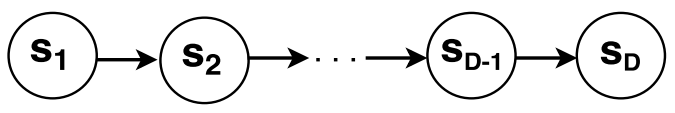

FIG. 4. *Directed Path Graph with D Nodes.*

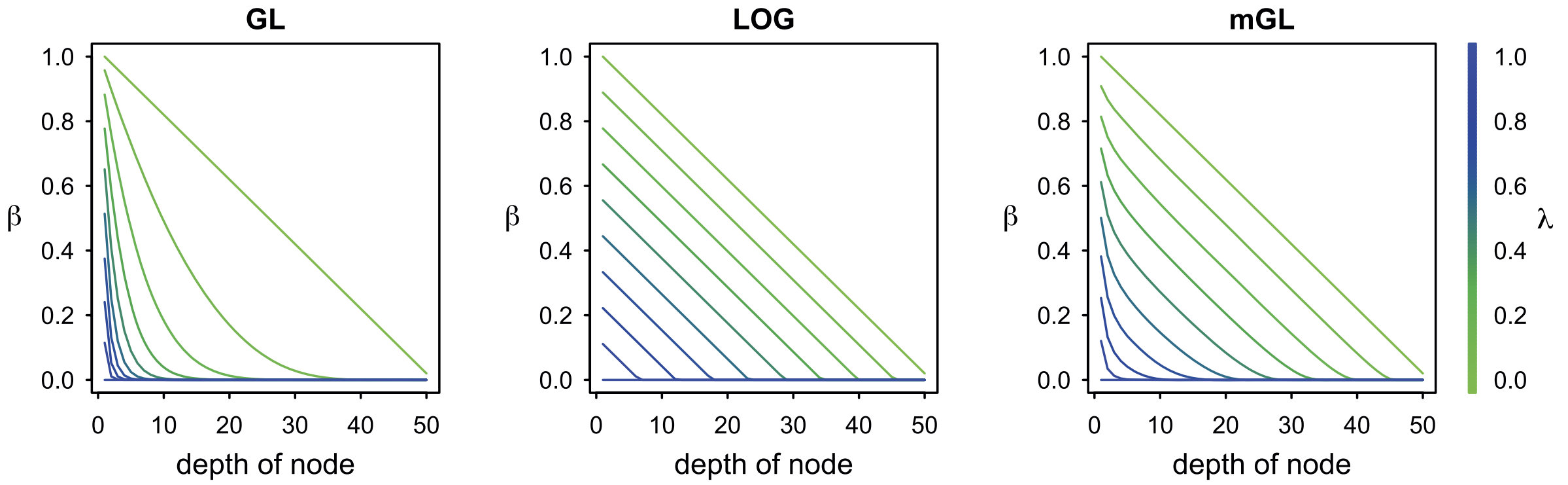

FIG. 5. *The effect of the proximal operator of three regularizers on* $\beta_i^* = 1 - \frac{i-1}{D}$: *Left*: $\hat\beta^{\mathrm{GL}}$, *middle*: $\hat\beta^{\mathrm{LOG}}$, *and right*: $\hat\beta^{\mathrm{mGL}}$.

for $i = 1, \dots, D$. For $\lambda \geq 0$, we compute

$$\hat\beta^{\mathrm{GL}} = \mathrm{Prox}_{\mathrm{GL}}^{d(\mathcal{D})}(y; \lambda, \{w_i = 1\}). \tag{11}$$

Likewise, the group structure $a(\mathcal{D})$ for LOG consists of groups of the form $\{1, \dots, i\}$ for $i = 1, \dots, D$, and we compute

$$\hat\beta^{\mathrm{LOG}} = \mathrm{Prox}_{\mathrm{LOG}}^{a(\mathcal{D})}(y; \lambda, \{w_i = \sqrt{i}\}). \tag{12}$$

The following two propositions emphasize the difference between the penalties in terms of the "overpenalization" phenomenon.

PROPOSITION 1. *Let* $\beta_d^* = 1_{\{d \leq K^*\}}$ *for* $K^* < D$. *For* $\hat\beta^{\mathrm{GL}}$ *in* (11), *if we choose* $\lambda > \bar\lambda := 2\sigma\sqrt{\log D}$, *then with probability at least* $1 - 2/D$,

(a) $\mathrm{supp}(\hat\beta^{\mathrm{GL}}) \subseteq \mathrm{supp}(\beta^*)$,
(b) *for* $1 \leq d \leq d + h \leq K^*$ *and* $\hat\beta_d^{\mathrm{GL}} \neq 0$,

$$\frac{|\hat\beta_{d+h}^{\mathrm{GL}}|}{|\hat\beta_d^{\mathrm{GL}}|} \leq \frac{|y_{d+h}|}{|y_d|} \exp\left(-\frac{\lambda h}{\sqrt{\sum_{m=d+1}^{K^*} y_m^2}}\right). \tag{13}$$

PROOF. See Appendix B.1. □

Equation (13) shows that the difference in the amount of shrinkage applied to two elements in $\mathcal{D}$ increases at least exponentially with the distance $h$ between them. In particular, Proposition 1 illustrates the differential shrinkage of GL: parameters embedded in nodes deep in the DAG are shrunken more aggressively than those that are in less deep nodes. Indeed, we can see this exponential decaying pattern empirically in two examples shown in the left panels of Figure 5 and Figure 6. The next proposition shows that LOG by contrast applies a uniform shrinkage across all elements.

PROPOSITION 2. *For the same* $\beta^*$ *as in Proposition* 1 *and* $\hat\beta^{\mathrm{LOG}}$ *in* (12), *assuming* $D > 1$ *and* $\bar\lambda := 2\sigma\sqrt{\log D} < 1$, *if we choose*

$$\bar\lambda < \lambda \leq (1 - \delta)(1 - \bar\lambda),$$

*for* $\delta \in (0, 1)$ *then with probability at least* $1 - 2/D$,

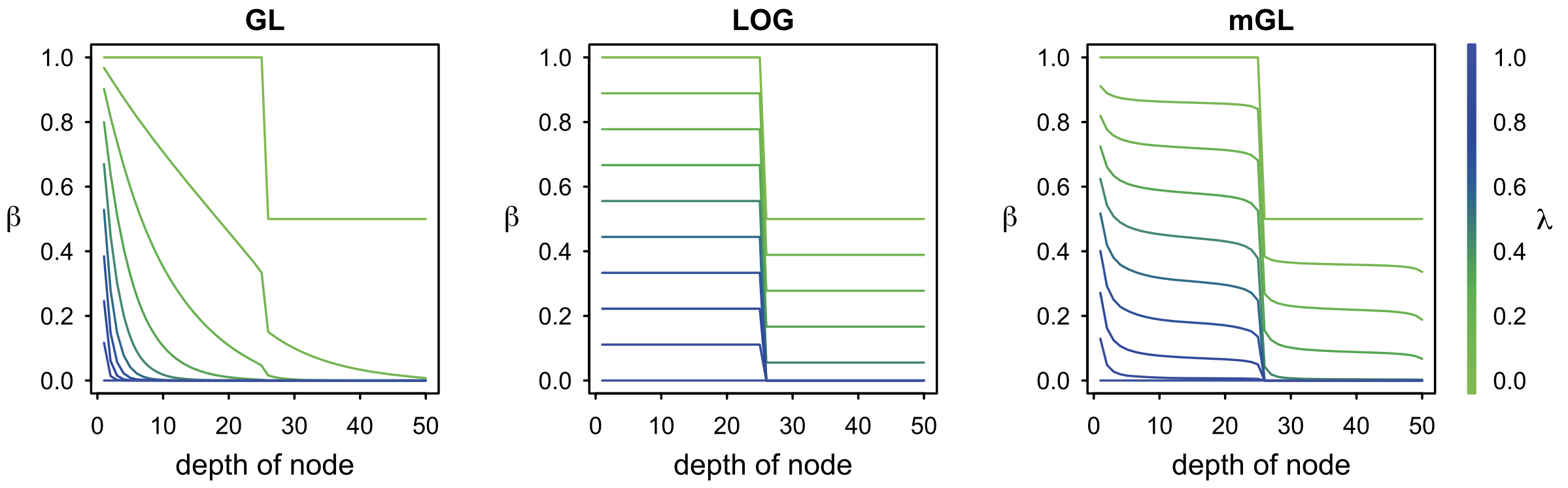

FIG. 6. *The effect of the proximal operator of three regularizers on* $\beta_i^* = 1_{\{i \leq D/2\}} + 0.5 * 1_{\{i > D/2\}}$: *Left*: $\hat\beta^{\mathrm{GL}}$, *middle*: $\hat\beta^{\mathrm{LOG}}$, *and right*: $\hat\beta^{\mathrm{mGL}}$.

(a) $\operatorname{supp}(\hat{\beta}^{\text{LOG}}) \subseteq \operatorname{supp}(\beta^*)$,
(b) *for* $1 \le d \le d+h \le K^*$ *and* $\hat{\beta}^{\text{LOG}}_{d+h} \neq 0$,

$$\delta \frac{|y_{d+h}|}{|y_d|} \le \frac{|\hat{\beta}^{\text{LOG}}_{d+h}|}{|\hat{\beta}^{\text{LOG}}_{d}|} \le \frac{|y_{d+h}|}{|y_d|}. \tag{14}$$

PROOF. See Appendix B.2. □

Equation (14) illustrates that the difference in the amount of shrinkage applied by LOG to two elements of different depths does not increase exponentially with the distance $h$ between the two elements. Moreover, the discrepancy in the amount of shrinkage is lower-bounded by a fixed quantity (that, importantly, does not depend on $h$) with high probability. For a fixed $\delta$, the range of $\lambda$ for which this holds is non-empty as $\sigma\sqrt{\log D} \to 0$. Proposition 2 thus establishes that LOG applies a comparable amount of shrinkage at all depths of $\mathcal{D}$. This is corroborated empirically in the middle panels of Figure 5 and Figure 6.

To demonstrate how pronounced the differential shrinkage phenomenon of GL is when the DAG depth is large, we plot the elements of $\hat{\beta}^{\text{GL}}$ and $\hat{\beta}^{\text{LOG}}$ when the depth is 50 (Figure 4 with $D = 50$). In order to better observe the effect of the proximal operator and thereby better understand the regularizer's influence, we consider a noiseless simulation, that is, $\sigma = 0$, and therefore $y = \beta^*$. We begin with a situation in which the input to the prox function decays linearly with depth, which might suggest to a statistician good reason to use a regularizer that shrinks elements deep in $\mathcal{D}$ to zero before others:

$$\beta_i^* = 1 - \frac{i-1}{D} \quad \text{for } i = 1, \dots, D.$$

The left and middle panels of Figure 5 show the proximal operators' outputs for ten equally spaced values of $\lambda$ between 0 and 1. When $\lambda$ is 0 (shown in green), both $\hat{\beta}^{\text{GL}}$ (in the left panel) and $\hat{\beta}^{\text{LOG}}$ (in the middle panel) simply return $y$. As we increase $\lambda$ (shown with increasing levels of blue), one notices a striking difference between the two regularizers. The LOG regularizer preserves the linear nature of the input while the GL regularizer shrinks elements deep in $\mathcal{D}$ to zero at a faster rate than those higher in $\mathcal{D}$. The result is that GL exaggerates the original downward trend in the input.

To balance the aggressive shrinkage of parameters appearing in many groups in the overlapping case, Jenatton, Audibert and Bach (2011) suggest weighting each parameter in a group differently based on the degree of overlaps existing on the parameter, instead of assigning a single weight to the whole group. In the context of banded covariance estimation, Bien, Bunea and Xiao (2016) also find that a better rate of convergence can be obtained using a more elaborate weighting scheme. For a fixed group $g_\ell \in d(\mathcal{D})$, the idea is to apply smaller weights to elements deeper in $\mathcal{D}$. In the directed path graph example, the weight applied to $s_m$ in group $g_\ell = \bigcup_{m=\ell}^{D} s_m$ is

$$w_{\ell,m} = \frac{1}{m - \ell + 1}, \quad \text{for } 1 \le \ell \le m \le D, \tag{15}$$

whereas a more general definition of the weights can be found in Appendix E.2. The modified GL (mGL) penalty and the corresponding proximal operator under the general weighting scheme can be denoted as

$$\Omega^{d(\mathcal{D})}_{\text{mGL}}(\beta; \{w_{\ell,m}\}) = \sum_{\ell=1}^{D} \sqrt{\sum_{m=\ell}^{D} w_{\ell,m}^2 \beta_m^2}, \tag{16}$$

$$\hat{\beta}^{\text{mGL}} = \operatorname*{arg\,min}_{\beta \in \mathbb{R}^p} \left\{ \frac{1}{2}\|y - \beta\|_2^2 + \lambda \Omega^{d(\mathcal{D})}_{\text{mGL}}(\beta; \{w_{\ell,m}\}) \right\}. \tag{17}$$

In the right panel of Figure 5, we see that $\hat{\beta}^{\text{mGL}}$ behaves less aggressively in shrinking elements deep in $\mathcal{D}$. In fact, it appears that GL with general weights mimics the LOG penalty.

Our second example considers a situation in which the raw input is a step function. We take $\beta_i^* = 1_{\{i \le D/2\}} + 0.5 * 1_{\{i > D/2\}}$ for $i = 1, \dots, D$. Figure 6 shows the effects of the three penalties. We find again that GL creates a strong downward trend whereas LOG preserves the relative sizes of the elements. Again, mGL behaves as a compromise between these two.

In summary, we observe that GL shrinks elements deep in $\mathcal{D}$ more than those high in $\mathcal{D}$. LOG by contrast is able to enforce the HSM constraints without applying differential shrinkage across $\mathcal{D}$. The mGL weighting scheme can effectively balance the aggressiveness of GL and seems reasonable to be used when more aggressive shrinkage is not desired. From a computational standpoint, which is the focus of the next section, this more elaborate weight structure complicates the computation of the proximal operator. Meanwhile, in some cases when the true model is sufficiently sparse, the GL approach, which favors simpler models, may serve a better role. Users should be aware of the difference among these frameworks and consequences, and choose a suitable approach based on their applications.

## 4. COMPUTATION

Given that both $\Omega_{\text{GL}}^{d(\mathcal{D})}$ and $\Omega_{\text{LOG}}^{a(\mathcal{D})}$ can be used in HSM, we would like to compare them from a computational perspective. Problems (6) and (8) are nonsmooth convex optimization problems, and proximal gradient methods (Nesterov, 2013, Beck and Teboulle, 2009) are well suited to such problems, especially when the non-differentiable part's proximal operator can be efficiently evaluated. We suppose that $F$ is differentiable and that $\nabla F$ is Lipschitz-continuous with constant $L$. In its simplest form, the proximal gradient method iteratively computes (for $k = 0, 1, 2, \ldots$)

$$\beta^{k+1} \leftarrow \arg\min_{\beta \in \mathbb{R}^p} \left\{ \frac{1}{2} \left\| \beta - \left( \beta^k - \frac{1}{L} \nabla F(\beta^k) \right) \right\|_2^2 + \lambda \Omega(\beta) \right\},$$

where $\Omega$ can be $\Omega_{\text{GL}}^{d(\mathcal{D})}$ or $\Omega_{\text{LOG}}^{a(\mathcal{D})}$. In words, at each step of the algorithm, the standard gradient descent step for minimizing $F$ is modified by applying the penalty $\lambda\Omega$'s proximal operator. It follows that an important computational benchmark lies in how efficiently the proximal operators, defined in (9) and (10), can be solved.

The proximal operator of GL when there are overlapping groups is usually solved via the dual problem (Boyd and Vandenberghe, 2004). As described in Jenatton et al. (2011), a dual of the proximal operator of (1) is given by

$$\min_{\{\eta^{(g)} \in \mathbb{R}^p\}_{g \in \mathcal{G}}} \left\{ \frac{1}{2} \left\| y - \sum_{g \in \mathcal{G}} \eta^{(g)} \right\|_2^2 \quad \text{s.t. } \|\eta^{(g)}\|_2 \le \lambda w_g \text{ and } \eta_{g^c}^{(g)} = 0 \text{ for } g \in \mathcal{G} \right\}.$$

Given a solution $\{\hat{\eta}^{(g)}\}_{g \in \mathcal{G}}$, it can be shown that $\text{Prox}_{\text{GL}}^{\mathcal{G}}(y; \lambda, w) = y - \sum_{g \in \mathcal{G}} \hat{\eta}^{(g)}$. The separable structure of the constraints suggests using block coordinate descent (BCD, Tseng, 2001) to solve for $\{\hat{\eta}^{(g)}\}_{g \in \mathcal{G}}$. Algorithm 1 has the details of implementation.

In the special case that $\mathcal{G} = d(\mathcal{D})$ and $\mathcal{D}$ is a tree, Jenatton et al. (2011) proves the remarkable result that the `while` loop in Algorithm 1 will terminate in one pass, as long as the pass of BCD over $g \in d(\mathcal{D})$ proceeds from innermost groups outward (i.e., from children to parents). The implication of this result is that when $\mathcal{D}$ is a tree, the proximal operator is essentially available in a closed form. Its computational complexity in this situation is $O(p)$, where $p$ is the dimension of $\beta$. By contrast, there is no known algorithm that solves the proximal operator of $\Omega_{\text{LOG}}^{a(\mathcal{D})}$ in a closed form under a tree structure. Several iterative methods have been used to solve (10), including cyclic projection (Villa et al., 2014) and BCD (Obozinski, Jacob and Vert, 2011). In Section 4.1, we review a commonly-used BCD approach for solving (10). In Section 4.2, we derive a new closed-form algorithm for solving (10) when $\mathcal{D}$ is a directed path graph. Finally, in Section 4.3, we leverage this new result to develop a more efficient algorithm for evaluating $\text{Prox}_{\text{LOG}}^{a(\mathcal{D})}$ for general DAGs $\mathcal{D}$.

**Algorithm 1** BCD in the dual for solving the proximal operator of $\Omega_{\text{GL}}^{\mathcal{G}}$

**Input:** $y, w, \lambda, \mathcal{G}$.
**Require:** $\lambda \ge 0$, $w_g > 0\ \forall g \in \mathcal{G}$.
$\eta^{(g)} = 0 \in \mathbb{R}^p$ for all $g \in \mathcal{G}$
$\beta = y$
**while** stopping criterion not reached **do**
  **for** $g \in \mathcal{G}$ **do**
    $\beta \leftarrow \beta + \eta^{(g)}$
    $\eta^{(g)} \leftarrow \frac{\lambda w_g \beta_g}{\|\beta_g\|_2}$
    $\beta \leftarrow \beta - \eta^{(g)}$
  **end for**
**end while**
**Output:** $\beta$

### 4.1 Naive BCD for LOG

By definition of the LOG penalty (2), its proximal problem can be rewritten in terms of the latent variables:

$$\min_{\beta \in \mathbb{R}^p} \left\{ \frac{1}{2} \|y - \beta\|_2^2 + \lambda \Omega_{\text{LOG}}^{\mathcal{G}}(\beta; w) \right\}$$

$$\Leftrightarrow \min_{\{v^{(g)} \in \mathbb{R}^p\}_{g \in \mathcal{G}}} \left\{ \frac{1}{2} \left\| y - \sum_{g \in \mathcal{G}} v^{(g)} \right\|_2^2 + \lambda \sum_{g \in \mathcal{G}} w_g \|v^{(g)}\|_2 \quad \text{s.t. } v_{g^c}^{(g)} = 0 \right\}.$$

In this parametrization, the penalty term naturally separates into blocks defined by the latent variables, and one can use BCD, cycling over the latent variable vectors (Obozinski, Jacob and Vert, 2011). Algorithm 2 provides the details of this approach, which we refer to as *naive BCD*.

The complexity per cycle of both Algorithm 1 and Algorithm 2 is $O(\sum_{g \in \mathcal{G}} |g|)$. Recalling that in HSM, for LOG, $\mathcal{G} = a(\mathcal{D})$ contains all ancestor sets whereas

**Algorithm 2** Naive BCD for solving the proximal operator of $\Omega_{\text{LOG}}^{\mathcal{G}}$

**Input:** $y, w, \lambda, \mathcal{G}$.
**Require:** $\lambda \geq 0$, $w_g > 0 \; \forall g \in \mathcal{G}$.
$v^{(g)} = 0 \in \mathbb{R}^p$ for all $g \in \mathcal{G}$
$\beta = 0 \in \mathbb{R}^p$
**while** stopping criterion not reached **do**
  **for** $g \in \mathcal{G}$ **do**
    $\beta \leftarrow \beta - v^{(g)}$
    $v^{(g)} \leftarrow S_G(y_g - \beta_g, \lambda w_g)$
    $\beta \leftarrow \beta + v^{(g)}$
  **end for**
**end while**
**Output:** $\beta$

for GL, $\mathcal{G} = d(\mathcal{D})$ contains all descendant sets. It is straightforward to observe that $a(\mathcal{D})$ and $d(\mathcal{D})$ have equal numbers of nodes in total. Assuming $|s_i|$ has the same magnitude across $i = 1, \ldots, N$, we see Algorithm 1 and Algorithm 2 require the same order of computation per cycle for general DAGs $\mathcal{D}$.

In the next section, we focus on the case in which $\mathcal{D}$ is a directed path graph and present a new algorithm that exactly solves the proximal operator in a finite number of steps. This will allow us to develop a more efficient alternative to naive BCD for general DAGs.

### 4.2 Closed-Form Solution of the LOG Prox for a Directed Path Graph

Suppose that $\mathcal{D}$ is a directed path graph with $D$ nodes as shown in Figure 4. We present here what can be seen as LOG counterpart to the result of Jenatton et al. (2011) for GL when $\mathcal{D}$ is a tree. For notational simplicity, we let $s_{i:j}$ denote $\bigcup_{k=i}^{j} s_k$. Using this notation, the group structure for the LOG penalty $a(\mathcal{D}) = \{s_{1:\ell} : \ell = 1, \ldots, D\}$ (since $s_{1:\ell}$ is the union of all indices contained in $s_i$ that are ancestors of $s_\ell$). A key quantity in Algorithm 3 is

$$f(j,k) = \frac{\|y_{s_{(k+1):j}}\|_2}{\sqrt{w_j^2 - w_k^2}}, \quad \text{for } 0 \leq k < j \leq D.$$

A standard choice for $w_j$ is $|s_{1:j}|^{1/2}$ in which case the denominator becomes $|s_{(k+1):j}|^{1/2}$ and $f(j,k)^2$ can be thought of as the average of $y_\ell^2$ for $\ell \in s_{(k+1):j}$. The algorithm identifies a sequence of knots $0 = k_0 < k_1 < \cdots < k_m \leq D$ with the properties that $k_i$ maximizes $f(\cdot, k_{i-1})$ and that $f(k_i, k_{i-1}) > \lambda$ for $i = 1, \ldots, m$.

**Algorithm 3** Solve the proximal operator of $\Omega_{\text{LOG}}^{a(\mathcal{D})}$ for a directed path graph $\mathcal{D}$

**Input:** $\lambda \geq 0$, $w = (w_1, \ldots, w_D) \in \mathbb{R}^{+D}$, $y \in \mathbb{R}^p$ and $a(\mathcal{D})$.
**Require:** $w_1 < \cdots < w_D$. $\mathcal{D}$ a path of depth $D$.
1: $\beta \leftarrow 0 \in \mathbb{R}^p$
2: $k \leftarrow 0 \in \mathbb{R}$
3: $w_0 \leftarrow 0 \in \mathbb{R}$
4: **while** $k < D$ **do**
5:   $K \leftarrow \arg\max_{j:j>k} f(j,k)$
6:   **if** $f(K,k) \leq \lambda$ **then**
7:     **break**
8:   **end if**
9:   $\beta_{s_{(k+1):K}} \leftarrow S_G(y_{s_{(k+1):K}}, \lambda\sqrt{w_K^2 - w_k^2})$
10:   $k \leftarrow K$
11: **end while**
**Output:** $\beta$

The knots are the values that $k$ has taken in the algorithm. Interestingly, once the set of knots has been determined, the algorithm is identical to that of the proximal operator of the non-overlapping group lasso with group structure $\{s_{(k_{i-1}+1):k_i}\}_{i=1,\ldots,m} \cup \{s_{1:D} \setminus s_{1:k_m}\}$ and weights $\{\sqrt{w_{k_i}^2 - w_{k_{i-1}}^2}\}_{i=1,\ldots,m} \cup \{\infty\}$. That is, each vector of elements between consecutive knots is separately groupwise soft-thresholded. The choice of knots implies that only the elements in $s_{1:D} \setminus s_{1:k_m}$ are set to zero. We see that the value of $\lambda$ determines the number of knots $m$, but not their location; thus, when solving the proximal operator for a sequence of $\lambda$ values, we only need to compute the knots once.

LEMMA 3. *Algorithm 3 computes the proximal operator in (10) for a directed path graph $\mathcal{D}$ of depth $D$ with complexity $O(p + Dm)$, where $m$ is the number of knots determined by the algorithm (not counting the initialization of $k = 0$). In the worst case when there are $D$ knots (i.e., $k$ increases by one and the condition in line 6 is never satisfied), the complexity is $O(p + D^2)$.*

PROOF. Appendix C proves that the algorithm computes the proximal operator, and Appendix D proves that when the solution has $m$ knots, Algorithm 3 requires $O(p + Dm)$ operations. To attain this complexity, one does not compute the $f(j,k)$ directly as defined in line 5 of the algorithm but rather performs constant time updates to reduce overall computation. □

In Appendix E, we show that the computational complexity of computing $\text{Prox}_{\text{GL}}^{d(\mathcal{D})}$ for this same DAG is $O(p + D)$. This means that when $D$ is larger than $p^{1/2}$, computing GL's prox may be more efficient than computing LOG's prox. By contrast, the computational complexity of computing the proximal operator of the modified GL penalty is $O(p + D^2 \log(n))$, given $n$-digit precision is required in using Newton's method for root-finding.

### 4.3 Path-Based BCD and ADMM for LOG

In the previous section, we showed that when $\mathcal{D}$ is a directed path graph, (10) can be solved extremely efficiently. For a general DAG $\mathcal{D}$, we can exploit this result by partitioning $\mathcal{D}$ into paths and cycling over the paths until convergence. The left panel of Figure 7 shows an example in which we partition a DAG into three paths. Let $\mathcal{P}_1, \dots, \mathcal{P}_L$ be our path decomposition of $\mathcal{D}$. We require that every node in $\mathcal{D}$ belongs to a unique path $\mathcal{P}_\ell$ and that the edges in path $\mathcal{P}_\ell$ all be in $\mathcal{D}$. The path decomposition of $\mathcal{D}$ induces a partition of $a(\mathcal{D})$ into $\mathcal{G}_1, \dots, \mathcal{G}_L$, where

$$\mathcal{G}_\ell = \{\text{ancestors}(\mathcal{D}; s_i) : s_i \in \mathcal{P}_\ell\}, \quad \text{for } \ell = 1, \dots, L.$$

The following lemma shows that the LOG penalty for a general DAG can be decomposed into a sum of LOG penalties, each having the simple path structure. This observation can be exploited to suggest an efficient alternative to naive BCD such that the "blocks" in the new approach are defined by the paths.

LEMMA 4. *Let $\{\mathcal{G}_\ell\}_{\ell=1}^L$ be the partition of $a(\mathcal{D})$ induced by the path decomposition $\mathcal{P}_1, \dots, \mathcal{P}_L$ of $\mathcal{D}$. For a convex smooth loss function $F(\beta)$, Problem (8) can be equivalently solved with*

(18)
$$\min_{\{\beta^{(\ell)} \in \mathbb{R}^p\}_{\ell=1}^L} \left\{ F\left(\sum_{\ell=1}^L \beta^{(\ell)}\right) + \lambda \sum_{\ell=1}^L \Omega_{\text{LOG}}^{\mathcal{G}_\ell}(\beta^{(\ell)}; w_{\mathcal{P}_\ell}) \quad s.t.\ \text{supp}(\beta^{(\ell)}) \subseteq \bigcup_{g \in \mathcal{G}_\ell} g \right\},$$

*where $w_{\mathcal{P}_\ell} = \{w_g : g \in \mathcal{G}_\ell\}$ for $\ell = 1, \dots, L$.*

PROOF. See Appendix F. □

Problem (18) satisfies the necessary conditions for BCD on $\beta^{(\ell)}$ to converge (Tseng, 2001). For solving the proximal Problem (10) where $F(\beta) = \frac{1}{2}\|y - \beta\|_2^2$, Algorithm 4 presents what we call *path-based BCD*. The value of this reparametrization is that each block update can be efficiently solved using Algorithm 3. When there are long paths in $\mathcal{D}$, the path-based BCD can make much faster progress compared to naive BCD since we are able to jointly minimize over all nodes in the path rather than settle for slow incremental progress. The decomposition of a DAG into paths is non-unique and the choice of path decomposition will affect efficiency. Algorithm 6 in Appendix G presents a simple greedy approach that attempts to break $\mathcal{D}$ into long paths. The *path-based BCD* is implemented in the R package `hsm` that accompanies this paper.

Clearly, the greatest efficiency gains for path-based BCD are to be expected when $\mathcal{D}$ can be decomposed into a small number of long path graphs. By contrast, the least favorable case for the path-based BCD is

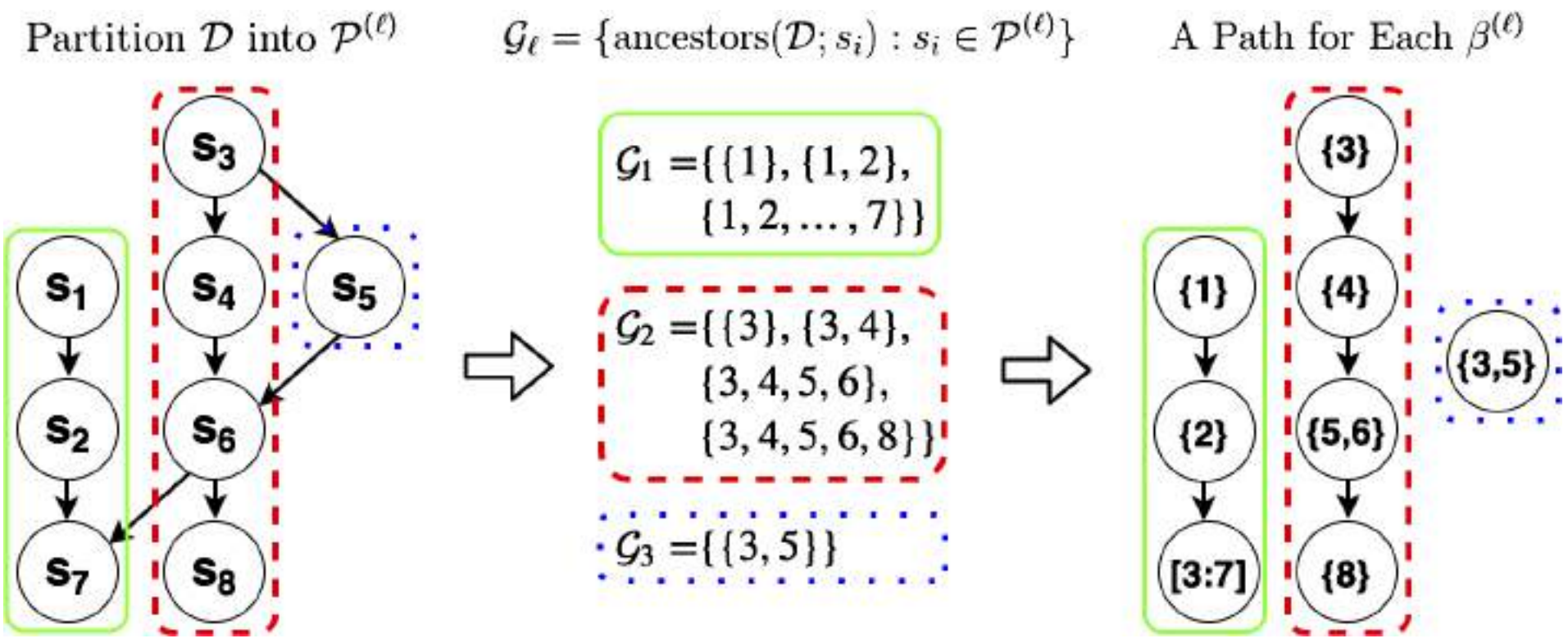

FIG. 7. *Let $s_i = \{i\}$ for $i \in \{1, \dots, 8\}$. Left: $a(\mathcal{D})$ is decomposed into 3 path graphs: $\mathcal{P}^{(1)}$ (in green solid contour), $\mathcal{P}^{(2)}$ (in red dashed contour) and $\mathcal{P}^{(3)}$ (in blue dotted contour). Middle: The partition of $\mathcal{G} = a(\mathcal{D})$: $\mathcal{G}_1$, $\mathcal{G}_2$ and $\mathcal{G}_3$ (colored accordingly). Right: $a(\mathcal{D})$ can be thought of as three separate path graphs on a new set of nodes, with parameter assignments shown inside each node*: $\text{supp}(\beta^{(1)}) \subseteq \{1, \dots, 7\}$ (*in green solid contour*), $\text{supp}(\beta^{(2)}) \subseteq \{3, 4, 5, 6, 8\}$ (*in red dashed contour*) *and* $\text{supp}(\beta^{(3)}) \subseteq \{3, 5\}$ (*in blue dotted contour*).

**Algorithm 4** Path-based BCD for solving the proximal operator of $\Omega_{\mathrm{LOG}}^{a(\mathcal{D})}$

**Input:** $y \in \mathbb{R}^p, w, \lambda, \mathcal{D}$, and a path-decomposition $\{\mathcal{P}_\ell\}_{\ell=1}^L$ of $\mathcal{D}$.

Generate $\mathcal{G}_\ell$ from $a(\mathcal{D})$ and $\{\mathcal{P}_\ell\}$.
$S_\ell \leftarrow \bigcup_{g\in\mathcal{G}_\ell} g$ for $\ell = 1, \dots, L$
$\beta^{(\ell)} \leftarrow 0 \in \mathbb{R}^p$ for $\ell = 1, \dots, L$
$\beta \leftarrow 0 \in \mathbb{R}^p$
**while** stopping criterion not reached **do**
  **for** $\ell \in [1 : L]$ **do**
    $\beta \leftarrow \beta - \beta^{(\ell)}$
    $\beta_{S_\ell}^{(\ell)} \leftarrow \mathrm{Prox}_{\mathrm{LOG}}^{\mathcal{G}_\ell}(y_{S_\ell} - \beta_{S_\ell}; \lambda, w_{\mathcal{P}_\ell})$ ▷ solved using Algorithm 3
    $\beta \leftarrow \beta + \beta^{(\ell)}$
  **end for**
**end while**

**Output:** $\beta$

when $\mathcal{D}$ is a depth-two tree since this structure does not have any long paths. The upper panel of Figure 8 shows these two trees along with a binary tree, which represents a choice for $\mathcal{D}$ between these two extremes. We perform simulations for these three choices of $\mathcal{D}$ to compare the rate of change of objective values using both BCD schemes. In the first example (upper left panel of Figure 8), $T_1$ and $T_2$ are path graphs of length 50 and 49, respectively, and each node has $|s_i| = 5$ (for a total of $p = 500$ parameters); in the second example (upper middle panel), we again have $|s_i| = 5$ (and $p = 500$); in the third example (upper right panel), we take a binary tree of depth 9, with $|s_i| = 1$ ($p = 2^9 - 1 = 511$). In all cases, we take $\lambda = 0.1$ and $w_g = |g|^{1/2}$.

For each $\mathcal{D}$, we randomly draw 20 samples of $y$ from $N_p(\mu = 0, \mathbf{\Sigma} = 4I_p)$, and use both methods to solve (10) at each $y$. The bottom panels of Figure 8 show the evolution over 50 cycles the ratio of the difference in objective values of the two BCDs and the difference in objective value of the path-based BCD and the "truth," evaluated at each cycle and averaged over 20 realizations. For each $(\mathcal{D}, y)$ pair, the objective evaluated at true parameter value is estimated with the minimum objective value computed over all the cycles of the two methods. All three curves are above zero after the starting point, indicating the naive approach is slower. In the most favorable case for path-based BCD (example 1), we see great advantage of using path-based BCD since the curve is in a much higher magnitude than the other two. As expected, path-based BCD has minor advantage over naive BCD in the depth-two tree case. For example, in the second cycle of the middle panel, the ratio takes value 0.8, meaning that (naive objective − true objective) is 80% larger than (path objective − true objective). In a non-extreme case represented by binary tree, path-based BCD still converges faster than naive BCD. For a more general $F(\beta) = \frac{1}{2}\|y - \mathbf{X}\beta\|_2^2$ in (8), Lemma 4 can be used to suggest an efficient alternating direction method of multipliers (ADMM, Boyd et al., 2011) approach:

LEMMA 5 (Path-based ADMM). *Let $\{\mathcal{G}_\ell\}_{\ell=1}^L$ be the partition of $a(\mathcal{D})$ induced by the path decomposi-*

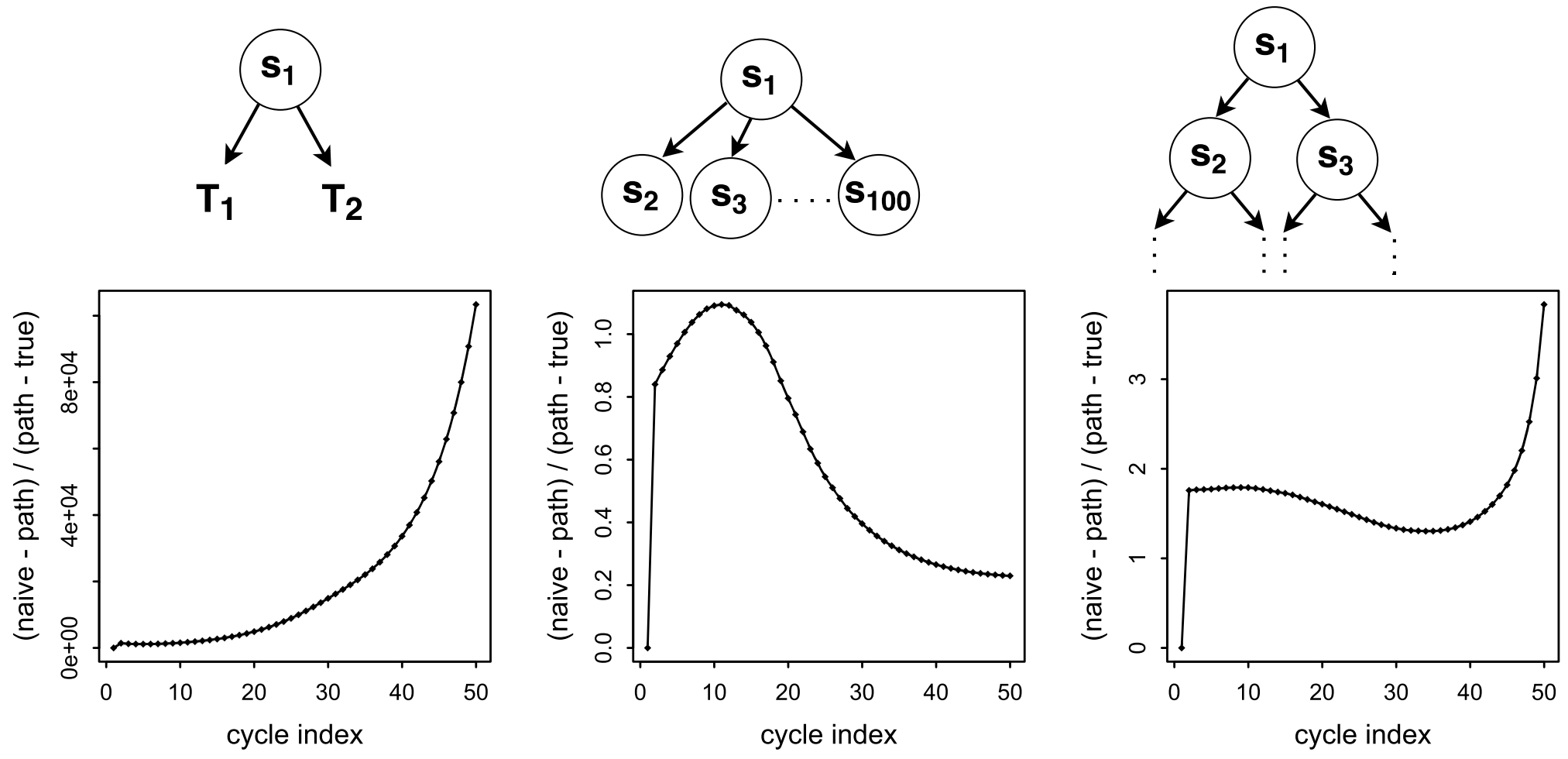

FIG. 8. *Top: Tree structures for example* 1, 2 *and* 3, *respectively. On top left, $T_1$ and $T_2$ are path graphs of length* 50 *and* 49, *respectively. Bottom: Plot of ratio of the difference in objective values of the two BCDs and the difference in objective value of the path-based BCD and the "truth", evaluated at each cycle and averaged over* 20 *realizations, with the corresponding tree above it.*

*tion $\mathcal{P}_1, \dots, \mathcal{P}_L$ of $\mathcal{D}$. For $y \in \mathbb{R}^n$ and $\mathbf{X} \in \mathbb{R}^{n\times p}$, Problem* (8) *with $F(\beta) = \frac{1}{2}\|y - \mathbf{X}\beta\|_2^2$ can be equivalently solved using ADMM on the following problem*:

(19)

$$\min_{\{\beta^{(\ell)}\in\mathbb{R}^p, \gamma^{(\ell)}\in\mathbb{R}^p\}_{\ell=1}^L} \frac{1}{2}\left\| y - \mathbf{X}\sum_{\ell=1}^L \gamma^{(\ell)}\right\|_2^2 + \lambda \sum_{\ell=1}^L \Omega_{\mathrm{LOG}}^{\mathcal{G}_\ell}(\beta^{(\ell)}; w_{\mathcal{P}_\ell})$$

$$s.t. \quad \beta^{(\ell)} = \gamma^{(\ell)} \text{ and } \operatorname{supp}(\beta^{(\ell)}) \subseteq \bigcup_{g\in\mathcal{G}_\ell} g =: g^{(\ell)} \quad \forall \ell = 1, \dots, L.$$

*The ADMM iterates among the following three steps and uses Algorithm* 4 *to solve Step* (2):

(1)

$$\hat{\gamma}^{(\ell)}_{|g^{(\ell)}} \leftarrow \hat{\beta}^{(\ell)}_{|g^{(\ell)}} + \frac{1}{\rho}\hat{u}^{(\ell)}_{|g^{(\ell)}} + \frac{1}{\rho}\mathbf{X}^T_{|g^{(\ell)}}(y - \Delta) \quad \forall \ell = 1, \dots, L$$

*where*

$$\Delta = \left(I + \frac{1}{\rho}\sum_\ell \mathbf{X}_{|g^{(\ell)}}\mathbf{X}^T_{|g^{(\ell)}}\right)^{-1} \cdot \sum_\ell\left(\mathbf{X}_{|g^{(\ell)}}\left(\hat{\beta}^{(\ell)}_{|g^{(\ell)}} + \frac{1}{\rho}\hat{u}^{(\ell)}_{|g^{(\ell)}}\right) + \frac{1}{\rho}\mathbf{X}_{|g^{(\ell)}}\mathbf{X}^T_{|g^{(\ell)}} y\right).$$

(2)

$$\hat{\beta}^{(\ell)}_{|g^{(\ell)}} \leftarrow \mathrm{Prox}_{\mathrm{LOG}}^{\mathcal{G}_\ell}\left(\left(\hat{\gamma}^{(\ell)}_{|g^{(\ell)}} - \frac{1}{\rho}\hat{u}^{(\ell)}_{|g^{(\ell)}}\right); \frac{\lambda}{\rho}, w_{\mathcal{P}_\ell}\right) \quad \forall \ell = 1, \dots, L.$$

(3)

$$\hat{u}^{(\ell)} \leftarrow \hat{u}^{(\ell)} + \rho(\hat{\gamma}^{(\ell)} - \hat{\beta}^{(\ell)}) \quad \forall \ell = 1, \dots, L.$$

PROOF. See Appendix H. □

## 5. ESTIMATING BANDED COVARIANCE WITH LOG

In Section 3, we observed that LOG avoids applying differential shrinkage on $\mathcal{D}$ as is in GL. In Section 4, we showed that when $\mathcal{D}$ is a directed path graph, the proximal operator can be evaluated in a closed form. In this section, we synthesize these observations in an application to covariance estimation. This example will demonstrate how choosing the LOG penalty leads to an estimator that achieves the statistical advantages of an existing estimator that requires the more complicated modified GL approach.

Suppose we observe a sample $X^{(1)}, X^{(2)}, \dots, X^{(n)} \in \mathbb{R}^p$ of independent zero-mean random vectors with true population covariance matrix $\mathbf{\Sigma}^*$. If the $p$ variables have a known ordering, a common assumption is that $\mathbf{\Sigma}^*$ is $K$-banded, meaning that

$$\mathbf{\Sigma}^*_{ij} = 0 \quad \text{for } |i - j| > K.$$

The sample covariance matrix, $\mathbf{S} = \frac{1}{n}\sum_{i=1}^n (X^{(i)} - \bar{X}) \cdot (X^{(i)} - \bar{X})^T$ (where $\bar{X} = \frac{1}{n}\sum_{i=1}^n X^{(i)}$), degrades as an estimator of $\mathbf{\Sigma}^*$ as $p$ increases; when $\mathbf{\Sigma}^*$ is (or could be reasonably approximated as) a banded matrix, banded estimators are preferable. It is straightforward to see that banded estimation of a matrix is an instance of HSM: Take $\mathcal{D}$ to be a directed path graph, such as that depicted in Figure 4, where

$$s_m = \{ij \in \{1, \dots, p\}^2 : |i - j| = m\} \quad \text{for } m = 1, \dots, p-1,$$

is the "subdiagonal" of elements that are $m$ away from the main diagonal. Bandedness of $\mathbf{\Sigma}$ can then be expressed as $\mathbf{\Sigma}_{s_\ell} = 0 \implies \mathbf{\Sigma}_{s_m} = 0$ for any $m > \ell$.

Bien, Bunea and Xiao (2016) propose "convex banding" estimators, which, in the terminology of our paper, correspond to

$$\hat{\mathbf{\Sigma}}^{\mathrm{GL}} = \operatorname*{arg\,min}_{\mathbf{\Sigma}\in\mathbb{R}^{p\times p}}\left\{\frac{1}{2}\|\mathbf{S} - \mathbf{\Sigma}\|_F^2 + \lambda\Omega_{\mathrm{GL}}^{d(\mathcal{D})}(\mathbf{\Sigma}^-; w)\right\} \quad \text{with } w_\ell = \sqrt{|s_\ell|}$$

being the weight on the group $s_{\ell:D}$, and

$$\hat{\mathbf{\Sigma}}^{\mathrm{mGL}} = \operatorname*{arg\,min}_{\mathbf{\Sigma}\in\mathbb{R}^{p\times p}}\left\{\frac{1}{2}\|\mathbf{S} - \mathbf{\Sigma}\|_F^2 + \lambda\Omega_{\mathrm{mGL}}^{d(\mathcal{D})}(\mathbf{\Sigma}^-; \tilde{w})\right\} \quad \text{with } \tilde{w}_{\ell,m} = \sqrt{|s_\ell|}/(m - \ell + 1)$$

being the weight on $s_m$ within the group $s_{\ell:D}$, where $\mathbf{\Sigma}^-$ denotes the matrix $\mathbf{\Sigma}$ but with zeros on its main diagonal. We recognize these as the proximal operators of the two penalties. Bien, Bunea and Xiao (2016) prove that both estimators can recover the true bandwidth with high probability; however, only $\hat{\mathbf{\Sigma}}^{\mathrm{mGL}}$, and not $\hat{\mathbf{\Sigma}}^{\mathrm{GL}}$, is shown to attain (up to a logarithmic factor) the minimax rate of convergence in Frobenius norm over a certain class of covariance matrices. They suggest, as we have here, that it is the overly aggressive shrinkage of subdiagonals far from the main diagonal (i.e., $s_m$ deep in $\mathcal{D}$) that prevents them from getting a similar rate for $\hat{\boldsymbol{\Sigma}}^{\text{GL}}$.

In light of our observation in Section 3 that LOG applies a comparable amount of shrinkage at all depths of $\mathcal{D}$, we investigate in this section whether a banded covariance estimator based instead on LOG can match the performance of $\hat{\boldsymbol{\Sigma}}^{\text{mGL}}$. Indeed, we will show that this LOG-based covariance estimator does successfully match the statistical performance of $\hat{\boldsymbol{\Sigma}}^{\text{mGL}}$, and, notably, does not require any modification of the weights as was the case with the GL-based estimator.

### 5.1 Defining the Estimator $\hat{\boldsymbol{\Sigma}}^{\text{LOG}}$

We define $\hat{\boldsymbol{\Sigma}}^{\text{LOG}}$ as the solution to the following problem:

$$\hat{\boldsymbol{\Sigma}}^{\text{LOG}} = \underset{\boldsymbol{\Sigma}\in\mathbb{R}^{p\times p}}{\arg\min}\left\{\frac{1}{2}\|\boldsymbol{\Sigma}-\mathbf{S}\|_F^2 + \lambda\Omega_{\text{LOG}}^{a(\mathcal{D})}(\boldsymbol{\Sigma}^-; w)\right\}, \tag{20}$$
$$\text{with } w_m = \sqrt{|s_{1:m}|}$$

being the weight on the group $s_{1:m}$. The group structure $a(\mathcal{D})$ is depicted in Figure 9. A key property of the "convex banding" estimators (Bien, Bunea and Xiao, 2016) is that they can be evaluated in a single pass over the elements of $\mathbf{S}$. By our result in Section 4.2, this advantageous computational property is shared by $\hat{\boldsymbol{\Sigma}}^{\text{LOG}}$. For completeness, Algorithm 3 in the context of covariance estimation is provided in Algorithm 7 of Appendix L.

### 5.2 Statistical Properties of $\hat{\boldsymbol{\Sigma}}^{\text{LOG}}$

We briefly review the statistical assumptions made in Bien, Bunea and Xiao (2016), which we will assume hold here as well.

ASSUMPTION 1. The random vector $X = (X_1, \dots, X_p)^T \in \mathbb{R}^p$ (which is mean 0 with covariance matrix $\boldsymbol{\Sigma}^*$) is marginally sub-Gaussian, that is,

$$\mathbb{E}\exp\big(tX_i/\sqrt{\boldsymbol{\Sigma}_{ii}^*}\big) \le \exp\big(Ct^2\big)$$

for all $t \ge 0$ and for some $C > 0$. Further, $\max_i |\boldsymbol{\Sigma}_{ii}^*| \le M$ for some constant $M > 0$.

ASSUMPTION 2. The dimension $p$ and sample size $n$ scale as follows: $\gamma_0 \log n \le \log p \le \gamma n$ for some $\gamma_0 > 0$, $\gamma > 0$.

Under these assumptions, it is proved in Lemma 1 of Bien, Bunea and Xiao (2016) that the random set

$$\mathcal{A}_x = \left\{\max_{1\le i,j\le p}\big|\mathbf{S}_{ij} - \boldsymbol{\Sigma}_{ij}^*\big| \le x\sqrt{\log p/n}\right\},$$

has high probability for sufficiently large $x$.

5.2.1 *Exact bandwidth recovery.* Suppose the true population covariance matrix $\boldsymbol{\Sigma}^*$ has bandwidth $K$, that is, we have $\boldsymbol{\Sigma}_{s_K}^* \ne 0$ and $\boldsymbol{\Sigma}_{s_k}^* = 0$ for $k > K$. Let $\hat{K}$ denote the bandwidth of $\hat{\boldsymbol{\Sigma}}^{\text{LOG}}$. We show in Theorem 1 and Theorem 2 that under mild conditions our estimator $\hat{\boldsymbol{\Sigma}}^{\text{LOG}}$ correctly recovers $K$ with high probability.

THEOREM 1. *If $\lambda \ge x\sqrt{\log p/n}$, then $\hat{K} \le K$ with high probability.*

PROOF. See Appendix I. □

From Theorem 1, we see that for large enough $\lambda$, $\hat{\boldsymbol{\Sigma}}^{\text{LOG}}$ will not overestimate $K$. In order for $\hat{\boldsymbol{\Sigma}}^{\text{LOG}}$ not to underestimate the true bandwidth, we need the nonzero elements of $\boldsymbol{\Sigma}^*$ to be sufficiently large. In the next theorem, we quantify the signal size by the root-mean-square of the elements of $\boldsymbol{\Sigma}^*$ in each group of the form $s_{m:K}$ for $m = 1, \dots, K$.

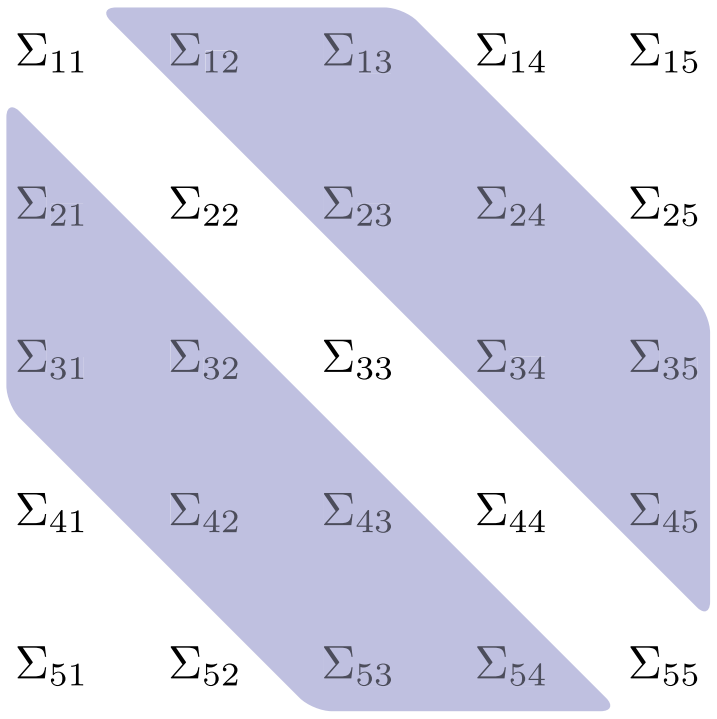

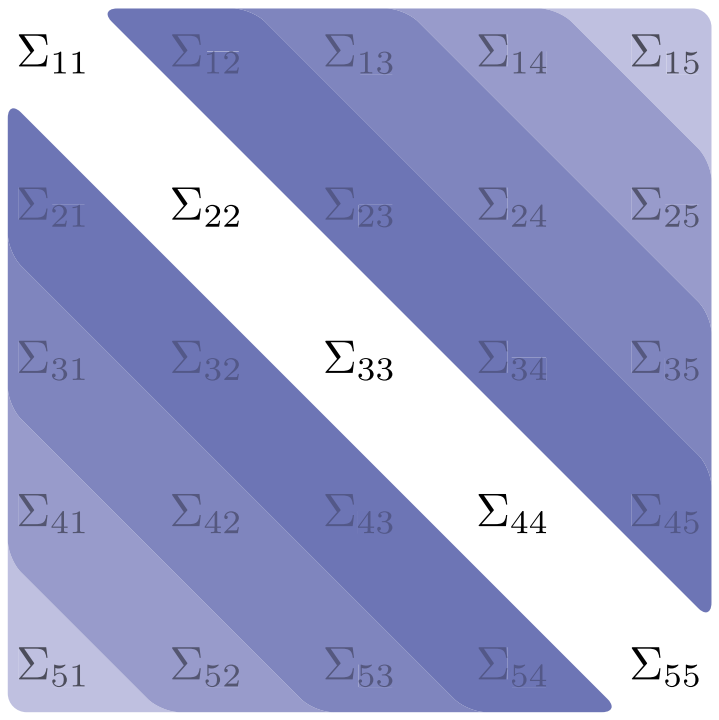

FIG. 9. *Left*: *The group $s_{1:2}$. Right*: *The nested groups of the form $s_{1:k}$ in $a(\mathcal{D})$.*

THEOREM 2. *Take* $\lambda$ *as in Theorem* 1. *If*

$$\min_{1\le m\le K} \frac{\|\boldsymbol{\Sigma}^*_{s_{m:K}}\|_F}{\sqrt{|s_{m:K}|}} > 2\lambda, \tag{21}$$

*then* $\hat{K} \ge K$ *with high probability.*

PROOF. See Appendix J. □

Thus, under the above signal strength condition, our LOG-based estimator correctly recovers the bandwidth with high probability. Furthermore, this condition is implied by the corresponding condition appearing in Theorem 4 of Bien, Bunea and Xiao (2016). This establishes that the LOG estimator recovers bandwidth at least as well as the "convex banding" estimators.

5.2.2 *Convergence in Frobenius norm.* In this section, we show that $\hat{\boldsymbol{\Sigma}}^{\text{LOG}}$ achieves, up to a multiplicative logarithmic factor, the optimal rate of convergence in Frobenius norm over the class of $K$-banded covariance matrices $\boldsymbol{\Sigma}^*$.

THEOREM 3. *Suppose* $\boldsymbol{\Sigma}^*$ *has bandwidth* $K$. *If* $\lambda = x\sqrt{\log p/n}$, *then with high probability*

$$\|\hat{\boldsymbol{\Sigma}}^{\text{LOG}} - \boldsymbol{\Sigma}^*\|_F^2 \lesssim \frac{pK\log p}{n}, \tag{22}$$

*where* $\lesssim$ *denotes an inequality holding up to a positive multiplicative constant independent of* $n$ *or* $p$.

PROOF. See Appendix K. □

This rate matches the statistical rate shown for $\hat{\boldsymbol{\Sigma}}^{\text{mGL}}$, but is noteworthy in that $\hat{\boldsymbol{\Sigma}}^{\text{LOG}}$ does not require the sophisticated weight structure of $\hat{\boldsymbol{\Sigma}}^{\text{mGL}}$.

### 5.3 Simulation Studies

From Section 5.2, we see that the estimators $\hat{\boldsymbol{\Sigma}}^{\text{LOG}}$ and $\hat{\boldsymbol{\Sigma}}^{\text{mGL}}$ have comparable theoretical properties; moreover, they both share the beneficial computational property that they can be computed in a single pass over the parameters. The more complicated weighting scheme of $\hat{\boldsymbol{\Sigma}}^{\text{mGL}}$ requires solving a one-dimensional line search for every subdiagonal whereas all operations in computing $\hat{\boldsymbol{\Sigma}}^{\text{LOG}}$ are very simple. We now further our comparison in two empirical studies. We consider two patterns for $\boldsymbol{\Sigma}^*$: a *moving-average pattern* and a *stair pattern.* The moving-average pattern corresponds to a downward linear decay in subdiagonal values:

$$\boldsymbol{\Sigma}^* = \text{toeplitz}\left(\left(1, \frac{K-1}{K}, \dots, \frac{1}{K}, 0_{p-K}\right)\right), \tag{23}$$

where toeplitz$(v)$ denotes a symmetric Toeplitz matrix with $v \in \mathbb{R}^p$ being the first column. The stair pattern, as its name suggests, adds flatness to the decay by introducing a "staircase" pattern in $\boldsymbol{\Sigma}^*$. We construct $\boldsymbol{\Delta} \in \mathbb{R}^{p\times p}$ as

$$\boldsymbol{\Delta} = \text{toeplitz}\big((1_{\frac{K}{5}}, 0.8 * 1_{\frac{K}{5}}, 0.6 * 1_{\frac{K}{5}}, 0.4 * 1_{\frac{K}{5}}, 0.2 * 1_{\frac{K}{5}}, 0_{p-K})\big)$$

and define

$$\boldsymbol{\Sigma}^* = \boldsymbol{\Delta} + \big(0.01 - \lambda_{\min}(\boldsymbol{\Delta})\big)_+ I_p \tag{24}$$

so that the minimum eigenvalue of $\boldsymbol{\Sigma}^*$ is at least 0.01.

For both studies, we simulate 50 samples of size 50 with a given $\boldsymbol{\Sigma}^*$, where each sample is denoted as $\{X^{(i)} \overset{\text{i.i.d.}}{\sim} N_p(0, \boldsymbol{\Sigma}^*)$ for $i = 1, \dots, 50\}$. A sample covariance $\mathbf{S}_j$ is computed with the $j$th sample. In terms of evaluating performance, we use *mean-squared error* as the metric of comparison:

$$\text{MSE}(\lambda) = \frac{1}{50}\sum_{j=1}^{50} \|\hat{\boldsymbol{\Sigma}}(\lambda, \mathbf{S}_j) - \boldsymbol{\Sigma}^*\|_F^2/p. \tag{25}$$

In the first study, we investigate to what extent the rate of $\hat{\boldsymbol{\Sigma}}^{\text{LOG}}$ derived in Theorem 3 in terms of $K$ and $p$ holds in practice. We simulate under the model used in Section 5.1.1 of Bien, Bunea and Xiao (2016). In particular, we take $\lambda_{\text{theory}} = 2\sqrt{\log p/n}$ and simulate with $\boldsymbol{\Sigma}^*$ in (23) for $p \in \{500, 1000, 2000\}$. At each $p$, we vary $K$ over 10 values equally spaced between 10 and 500. In agreement with Theorem 3, the left panel of Figure 10 shows (for three values of $p$) an approximate linear dependence of $K$ on squared Frobenius norm. The right panel supports the $p$ dependence of Theorem 3 since we find that the three curves line up when we scale the squared Frobenius norm by $p \log p$.

In the second study, we compare the empirical performance of $\hat{\boldsymbol{\Sigma}}^{\text{GL}}$, $\hat{\boldsymbol{\Sigma}}^{\text{mGL}}$, and $\hat{\boldsymbol{\Sigma}}^{\text{LOG}}$ over the two patterns for $\boldsymbol{\Sigma}^*$ at $p = 500$ and for various $K$. In contrast to the previous study, where we used the theoretically justified $\lambda_{\text{theory}}$ of the form $x\sqrt{\log p/n}$, in this study we use

$$\lambda_{\text{best}} = \arg\min_{\lambda\in\Lambda} \text{MSE}(\lambda), \tag{26}$$

where $\Lambda$ is a grid of 50 values equally spaced on the log scale. The quantity MSE$(\lambda_{\text{best}})$ is an estimate of $\min_\lambda \mathbb{E}\|\hat{\boldsymbol{\Sigma}}(\lambda) - \boldsymbol{\Sigma}^*\|_F^2/p$ and provides a view of the best obtainable performance of each method.

We first consider the moving-average pattern described in (23) for $\boldsymbol{\Sigma}^*$ with $K$ varying over 10 equally-spaced values between 10 and 500. The left panel of

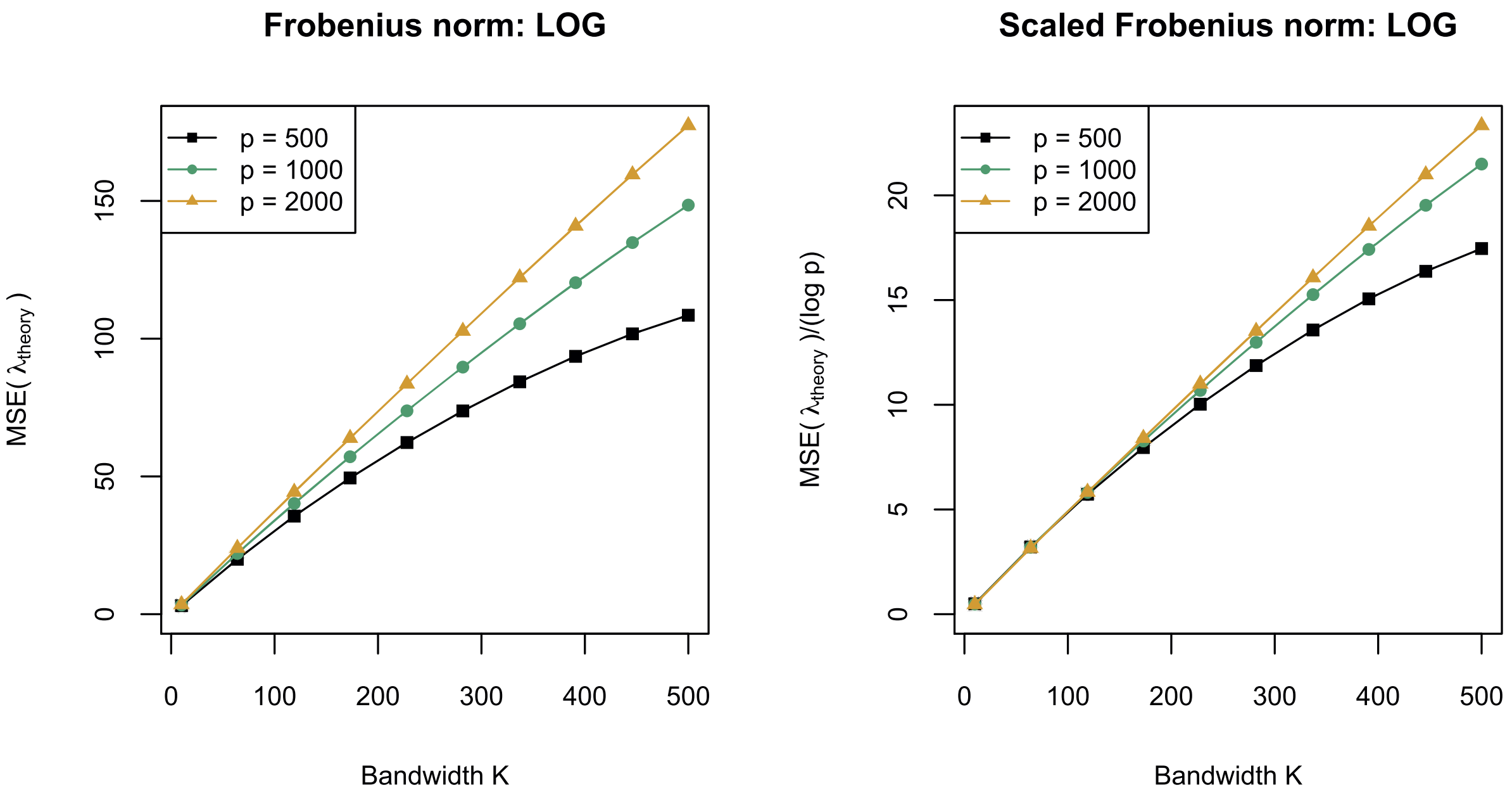

FIG. 10. *Left*: $\mathrm{MSE}(\lambda_{\text{theory}})$, *and right*: $\mathrm{MSE}(\lambda_{\text{theory}})/\log p$ *as a function of* $K$ *for* $\hat{\boldsymbol{\Sigma}}^{\text{LOG}}$ *where* $\lambda_{\text{theory}} = 2\sqrt{\log p/n}$.

Figure 11 shows how $\mathrm{MSE}(\lambda_{\text{best}})$ varies with $K$ for the three methods. We notice that $\hat{\boldsymbol{\Sigma}}^{\text{GL}}$ outperforms $\hat{\boldsymbol{\Sigma}}^{\text{mGL}}$ and $\hat{\boldsymbol{\Sigma}}^{\text{LOG}}$ at all $K$. In addition, $\hat{\boldsymbol{\Sigma}}^{\text{mGL}}$ and $\hat{\boldsymbol{\Sigma}}^{\text{LOG}}$ appear to perform similarly. It is striking to compare the scale of the $y$-axis in the left panel of Figure 10 to that of Figure 11. Figure 10 shows the performance of $\hat{\boldsymbol{\Sigma}}^{\text{LOG}}$ with $\lambda_{\text{theory}} = 2\sqrt{\log p/n}$, which while motivated by theory, is evidently far from the optimal choice of $\lambda$ in terms of MSE. The sublinear curve seen in Figure 11 is again a reminder that the theory is about $\lambda = x\sqrt{\log p/n}$ and not about $\lambda_{\text{best}}$.

The second pattern we consider for $\boldsymbol{\Sigma}^*$ is the stair pattern described in (24) with $K$ varying over 10 equally-spaced values between 50 and 500. As shown in the right panel of Figure 11, all three estimators achieve much larger error than in the moving average case. When $K$ is small ($K < 200$), $\hat{\boldsymbol{\Sigma}}^{\text{GL}}$ beats $\hat{\boldsymbol{\Sigma}}^{\text{mGL}}$ and $\hat{\boldsymbol{\Sigma}}^{\text{LOG}}$, but by a small amount. When $K$ becomes larger, both $\hat{\boldsymbol{\Sigma}}^{\text{mGL}}$ and $\hat{\boldsymbol{\Sigma}}^{\text{LOG}}$ outperform $\hat{\boldsymbol{\Sigma}}^{\text{GL}}$. We again see similar performance between $\hat{\boldsymbol{\Sigma}}^{\text{mGL}}$ and $\hat{\boldsymbol{\Sigma}}^{\text{LOG}}$. The relative performance of these three methods in these two scenarios suggests that LOG and mGL perform very similarly and that it is difficult to say in general whether these perform better or worse than GL.

Since we are estimating a covariance matrix, we are also interested in getting a positive semidefinite (PSD) estimate. For the stair pattern, we find in simulation that these three estimators are always PSD. By contrast, in the moving-average example, we find that the probability of each estimator being PSD at each method's $\lambda_{\text{best}}$ varies with $K$ (see Figure 12 of Appendix M). We find that the probability that $\hat{\boldsymbol{\Sigma}}^{\text{GL}}$ is PSD decreases to 0 as $K$ increases to $p$. For $\hat{\boldsymbol{\Sigma}}^{\text{mGL}}$ and $\hat{\boldsymbol{\Sigma}}^{\text{LOG}}$, the $K$ dependence is less simple; for large $K$, the probability that they are PSD is approximately 80%, but for moderate $K$, we find the probability drops to as low as 20%. If positive definiteness is important in a given application, one could modify Problem (20) to include a PSD constraint as is done in Problem (2.3) of Bien, Bunea and Xiao (2016).

## 6. CONCLUSION

In this paper, we focus on hierarchical sparse modeling, a structure that arises in a wide array of statistical problems. In particular, we investigate the differences between two convex penalties, GL and LOG, that have been used in this context for identical purposes but until now have not been systematically compared for HSM.

We highlight a phenomenon of GL in which parameters embedded deep within the HSM's DAG are more aggressively regularized than those that are less deeply embedded. We find that this phenomenon may have negative statistical consequences for GL—both theoretical and empirical—when the DAG has deep nodes and the true model is not very sparse. While a modification of GL is possible to curb this over-aggressiveness of GL (Jenatton, Audibert and Bach, 2011, Bach et al., 2012, Bien, Bunea and Xiao, 2016), doing so complicates the computation and makes for a more difficult to describe estimator. By contrast, we show that using LOG fulfills our goal without any additional complication and performs, both in practice and in theory, very similarly to the modified GL penalty. In the special case that the DAG is a path, we derive a closed-form expression for the proximal operator of LOG that can be seen as the LOG counterpart to a result of Jenatton et al. (2011) about the GL penalty. Having this closed form makes computation extremely efficient for directed path graphs, and we leverage this efficiency to general DAGs and more general problems by proposing path-based BCD and path-based ADMM algorithms. We show in simulation that the path-based BCD algorithm converges in fewer passes over the parameters than the standard BCD approach for LOG.

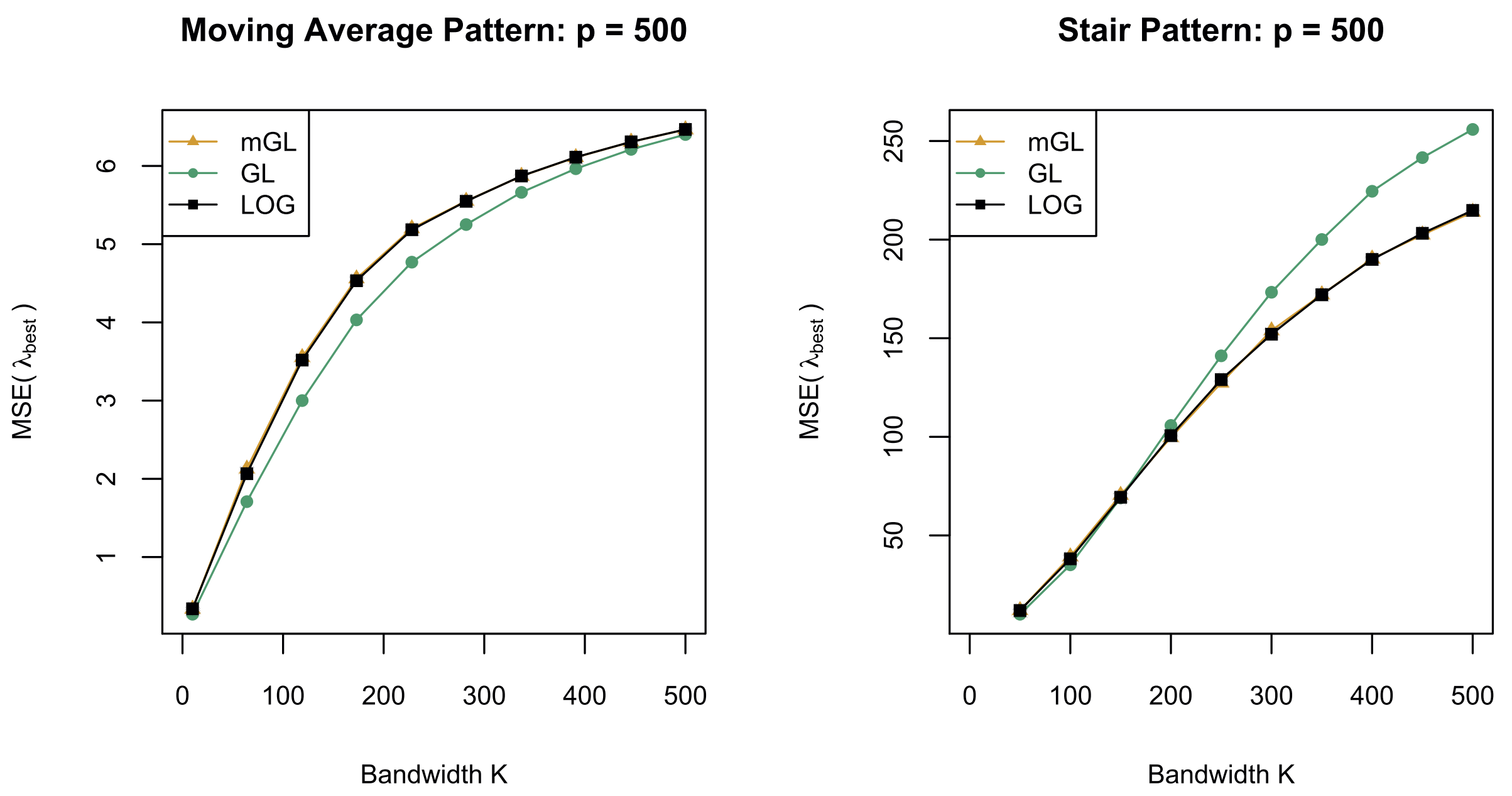

FIG. 11. *For the three estimators* ($\hat{\mathbf{\Sigma}}^{\text{mGL}}$, $\hat{\mathbf{\Sigma}}^{\text{GL}}$, $\hat{\mathbf{\Sigma}}^{\text{LOG}}$), $\text{MSE}(\lambda_{\text{best}})$ *as a function of* $K$ *under the moving average pattern* (*left*) *and the stair pattern* (*right*) *where* $\lambda_{\text{best}} = \arg\min_{\lambda\in\Lambda} \text{MSE}(\lambda)$.

As an application of these ideas to statistics, we show how the recent "convex banding" covariance estimator of Bien, Bunea and Xiao (2016) could have instead been formulated with an LOG penalty. We show that our LOG-based estimator attains the same convergence and recovery results as the mGL-based appoach in Bien, Bunea and Xiao (2016) and in simulation performs extremely similarly as well. The advantage of our LOG estimator is that it is easier to describe and compute.

In future work, it would be interesting to determine whether a closed-form solution exists for DAG structures more general than a directed path graph. While we were not able to derive such a closed form, we have not established that such a solution does not exist. Another avenue for future work lies in extending the comparison of GL and LOG to situations beyond the class of problems considered here. For example, the sparse group lasso penalty, $\sum_{k=1}^{K} w_k\|\beta_{g_k}\|_2 + \|\beta\|_1$ (Simon et al., 2013) is a GL penalty with $K + p$ groups: $g_1, \dots, g_K, \{1\}, \dots, \{p\}$. This group structure can be written as $d(\mathcal{D})$, where $\mathcal{D}$ is a forest of $K$ trees, each having an empty root pointing to the singletons contained in $g_k$. However, the LOG penalty on $a(\mathcal{D})$ is simply the lasso, whereas an LOG with $g_1, \dots, g_K, \{1\}, \dots, \{p\}$ would seem to be the appropriate corresponding model.

## APPENDIX A: PROOF OF LEMMA 1

For $p = 2$, denote $\beta = (\beta_1, \beta_2) \in \mathbb{R}^2$. The $\Omega_{\text{GL}}^{d(\mathcal{D})}$ and $\Omega_{\text{LOG}}^{a(\mathcal{D})}$ penalties can be written as

$$\Omega_{\text{GL}}^{d(\mathcal{D})}(\beta; w) = w_1\|(\beta_1, \beta_2)\|_2 + w_2|\beta_2|,$$

$$\Omega_{\mathrm{LOG}}^{a(\mathcal{D})}(\beta; w') = \min_{\{v_1^{(1)}\in\mathbb{R}, v^{(2)}\in\mathbb{R}^2\}} \left\{ |v_1^{(1)}| + \|v^{(2)}\|_2 \quad \text{s.t.} \begin{pmatrix} v_1^{(1)} + v_1^{(2)} \\ v_2^{(2)} \end{pmatrix} = \begin{pmatrix} \beta_1 \\ \beta_2 \end{pmatrix} \right\}.$$

Suppose there exists $w \in \mathbb{R}^{+2}$ such that for all $\beta$, $\Omega_{\mathrm{GL}}^{d(\mathcal{D})}(\beta; w) = \Omega_{\mathrm{LOG}}^{a(\mathcal{D})}(\beta; w')$ holds. The equality also holds for $\beta = (0, \beta_2)$ and $\beta = (\beta_1, 0)$.

- When $\beta = (0, \beta_2)$, that is, $\beta_1 = 0$, the following is true:

$$\Omega_{\mathrm{LOG}}^{a(\mathcal{D})}(\beta; w') = \min_{v_1^{(1)}\in\mathbb{R}} |v_1^{(1)}| + \sqrt{(v_1^{(1)})^2 + \beta_2^2} = |\beta_2| = \Omega_{\mathrm{GL}}^{d(\mathcal{D})}(\beta; w) = (w_1 + w_2)|\beta_2|.$$

  We get $w_1 + w_2 = 1$.
- When $\beta = (\beta_1, 0)$, that is, $\beta_2 = 0$, the following is true:

$$\Omega_{\mathrm{LOG}}^{a(\mathcal{D})}(\beta; w') = \min_{v_1^{(2)}\in\mathbb{R}} |\beta_1 - v_1^{(2)}| + |v_1^{(2)}| = |\beta_1| = \Omega_{\mathrm{GL}}^{d(\mathcal{D})}(\beta; w) = w_1|\beta_1|.$$

  We get $w_1 = 1$.

Combining the results above we have $w_2 = 0$ which leads to a contradiction. Hence, when $p = 2$ and $w' = (1, 1)$, there does not exist $w \in \mathbb{R}^{+2}$ such that $\Omega_{\mathrm{GL}}^{d(\mathcal{D})}(\cdot; w) = \Omega_{\mathrm{LOG}}^{a(\mathcal{D})}(\cdot; w')$.

## APPENDIX B: PROOF OF PROPOSITIONS 1 AND 2

Let $y = \beta^* + \varepsilon$ where $\varepsilon \sim N_D(0, \sigma^2 I_D)$ and $\beta_d^* = 1_{\{d\le K^*\}}$ for $d = 1, \dots, D$ (and assume $K^* < D$). We define the event

$$\mathcal{B} := \left\{ \max_{i=1,\dots,D} |\varepsilon_i| > \bar{\lambda} \right\}, \tag{27}$$

where $\bar{\lambda} := 2\sigma\sqrt{\log D}$. A union bound and a Chernoff upper bound for normal variables establishes that

$$\mathbb{P}\Big(\max_{i=1,\dots,D} |\varepsilon_i| > t\Big) \le D\mathbb{P}(|\varepsilon_1| > t) \le 2De^{-t^2/2\sigma^2},$$

for $t > 0$. Taking $t = 2\sigma\sqrt{\log D}$ gives that

$$\mathbb{P}(\mathcal{B}) \le 2/D. \tag{28}$$

### B.1 Proof of Proposition 1

We establish the following deterministic result that holds on $\mathcal{B}^c$ [by (28), this proves Proposition 1].

LEMMA 6. *The following two statements hold on $\mathcal{B}^c$ under the assumptions of Proposition 1:*

(a) $\mathrm{supp}(\hat{\beta}^{\mathrm{GL}}) \subseteq \mathrm{supp}(\beta^*)$,
(b) *for* $1 \le d \le d + h \le K^*$ *and* $\hat{\beta}_d^{\mathrm{GL}} \ne 0$,

$$\frac{|\hat{\beta}_{d+h}^{\mathrm{GL}}|}{|\hat{\beta}_d^{\mathrm{GL}}|} \le \frac{|y_{d+h}|}{|y_d|} \exp\left(-\frac{\lambda h}{\sqrt{\sum_{m=d+1}^{K^*} y_m^2}}\right).$$

PROOF. Jenatton et al. (2011) provide a closed-form solution for (9) (see Algorithm 2 in their paper). Their algorithm in this context is as follows:

1. Initialize $\hat{b}^{(D)} = y$.
2. For $d = D, \dots, 1$,

$$\hat{b}_{d:D}^{(d-1)} \leftarrow \hat{b}_{d:D}^{(d)} \cdot \left(1 - \frac{\lambda}{\|\hat{b}_{d:D}^{(d)}\|_2}\right)_+, \quad \text{and}$$

$$\hat{b}_{1:(d-1)}^{(d-1)} \leftarrow y_{1:(d-1)} \quad \text{if } d > 1.$$

Defining $\hat{r}_d := \|\hat{b}_{d:D}^{(d)}\|_2$ for $d = 1, \dots, D$, one gets the recurrence relation

$$\hat{r}_{d-1}^2 = (\hat{r}_d - \lambda)_+^2 + y_{d-1}^2 \quad \text{where } \hat{r}_D = |y_D|. \tag{29}$$

The solution to (9) can be expressed, for each $d$, as

$$\hat{\beta}_d^{\mathrm{GL}} = y_d \cdot \prod_{\ell=1}^{d} (1 - \lambda/\hat{r}_\ell)_+. \tag{30}$$

Our choice of $\lambda$ in Proposition 1 establishes that on $\mathcal{B}^c$, $\lambda > \max_{i=1,\dots,D} |\varepsilon_i|$. This together with the recurrence relation in (29) implies the following:

- For $d = K^* + 1, \dots, D$, $\hat{r}_d = |y_d| = |\varepsilon_d|$ and thus, by (30), $\hat{\beta}_d^{\mathrm{GL}} = 0$ for $d > K^*$. This establishes that $\mathrm{supp}(\hat{\beta}^{\mathrm{GL}}) \subseteq \mathrm{supp}(\beta^*)$.
- For $d = K^*$,

$$\hat{r}_{K^*} = |y_{K^*}| \le \sqrt{\sum_{\ell=d}^{K^*} y_\ell^2}.$$

  For $d < K^*$, suppose $\hat{r}_{d+1} \le \sqrt{\sum_{\ell=d+1}^{K^*} y_\ell^2}$. Then we have

$$\hat{r}_d = \sqrt{(\hat{r}_{d+1} - \lambda)_+^2 + y_d^2} \le \sqrt{\hat{r}_{d+1}^2 + y_d^2} \le \sqrt{\sum_{\ell=d+1}^{K^*} y_\ell^2 + y_d^2} \le \sqrt{\sum_{\ell=d}^{K^*} y_\ell^2}.$$

  This establishes by induction that $\hat{r}_d \le \sqrt{\sum_{\ell=d}^{K^*} y_\ell^2}$ for all $d \le K^*$.

For $1 \le d \le d+h \le K^*$, assuming $\hat{\beta}_d^{\mathrm{GL}} \neq 0$, we have

$$\begin{aligned}
\frac{|\hat{\beta}_{d+h}^{\mathrm{GL}}|}{|\hat{\beta}_d^{\mathrm{GL}}|} &= \frac{|y_{d+h}| \cdot \prod_{\ell=1}^{d+h}(1-\frac{\lambda}{\hat{r}_\ell})_+}{|y_d| \cdot \prod_{\ell=1}^{d}(1-\frac{\lambda}{\hat{r}_\ell})_+} \\
&= \frac{|y_{d+h}|}{|y_d|} \prod_{\ell=d+1}^{d+h}\left(1-\frac{\lambda}{\hat{r}_\ell}\right)_+ \\
&\le \frac{|y_{d+h}|}{|y_d|} \exp\left(-\sum_{\ell=d+1}^{d+h}\frac{\lambda}{\hat{r}_\ell}\right) \\
&\qquad \left(\text{since } (1-x)_+ \le e^{-x} \text{ for } x \in \mathbb{R}\right) \\
&\le \frac{|y_{d+h}|}{|y_d|} \exp\left(-\sum_{\ell=d+1}^{d+h}\frac{\lambda}{\sqrt{\sum_{m=\ell}^{K^*} y_m^2}}\right) \\
&\le \frac{|y_{d+h}|}{|y_d|} \exp\left(-\frac{\lambda h}{\sqrt{\sum_{m=d+1}^{K^*} y_m^2}}\right). \qquad \square
\end{aligned}$$

### B.2 Proof of Proposition 2

We prove two deterministic lemmas, corresponding to parts (a) and (b) in Proposition 2.

LEMMA 7. *Under the assumptions of Proposition 2*, $\mathrm{supp}(\hat{\beta}^{\mathrm{LOG}}) \subseteq \mathrm{supp}(\beta^*)$ *holds on* $\mathcal{B}^c$.

PROOF. We prove this using Algorithm 3 which solves (10) under a directed path graph. Let $\bar{K}$ be the largest knot such that $\bar{K} \le K^*$, determined by Algorithm 3 on solving (10). We show in what follows that $f(k, \bar{K}) \le \lambda \ \forall k > \bar{K}$, which establishes that $\bar{K}$ is the last knot. The assumed lower bound on $\lambda$ ensures that $\lambda > \max_{i=1,\dots,D} |\varepsilon_i|$ on the event $\mathcal{B}^c$.

If $\bar{K} = K^*$, $\forall k > K^*$

$$(31)\qquad \begin{aligned} f(k,\bar{K}) &= \frac{\|y_{(\bar{K}+1):k}\|_2}{\sqrt{k-\bar{K}}} = \frac{\|\varepsilon_{(K^*+1):k}\|_2}{\sqrt{k-K^*}} \\ &\le \max_{i=1,\dots,D} |\varepsilon_i| < \lambda. \end{aligned}$$

If $\bar{K} < K^*$, then $K^*$ is not chosen as a knot, given that $\bar{K}$ by construction is the last knot determined by Algorithm 3 on or before $K^*$. We let $k = \bar{K}$ on line 5 of Algorithm 3. Consider two possible cases for $K^*$: $K^* = \arg\max_{j:j>\bar{K}} f(j,\bar{K})$ and $K^* \neq \arg\max_{j:j>\bar{K}} f(j,\bar{K})$. In the first case, we must have $f(K^*,\bar{K}) \le \lambda$ otherwise the while loop would not break on line 7, making $K^*$ a knot and leading to a contradiction. In the second case, let $\check{K} := \arg\max_{j:j>\bar{K}} f(j,\bar{K})$. If $\check{K} < K^*$, we have $f(K^*,\bar{K}) \le f(\check{K},\bar{K}) \le \lambda$ otherwise $\check{K}$ would be a knot which would then be in contradiction with the assumption that $\bar{K}$ was the last knot on or before $K^*$. If $\check{K} > K^*$, we have $f(K^*,\bar{K}) \le f(\check{K},\bar{K})$ by definition of $\check{K}$. In summary, either one of the following is true for the second case:

1. $f(K^*,\bar{K}) \le \lambda$.
2. $\exists \bar{k} > K^*$ such that $f(K^*,\bar{K}) \le f(\bar{k},\bar{K})$, that is, $\|y_{(\bar{K}+1):K^*}\|_2^2 \le \|y_{(\bar{K}+1):\bar{k}}\|_2^2 \cdot \frac{K^*-\bar{K}}{\bar{k}-\bar{K}}$.

We show that in both cases

$$(32)\qquad \|y_{(\bar{K}+1):K^*}\|_2^2 \le \lambda^2(K^*-\bar{K}).$$

Case (i) is equivalent to (32). When Case (ii) holds, $\exists \bar{k} > K^*$ such that

$$(33)\qquad \begin{aligned} &\|y_{(\bar{K}+1):K^*}\|_2^2 \\ &\le \|y_{(\bar{K}+1):\bar{k}}\|_2^2 \cdot \frac{K^*-\bar{K}}{\bar{k}-\bar{K}} \\ &= \left(\|y_{(\bar{K}+1):K^*}\|_2^2 + \|\varepsilon_{(K^*+1):\bar{k}}\|_2^2\right) \cdot \frac{K^*-\bar{K}}{\bar{k}-\bar{K}}. \end{aligned}$$

Plugging $\alpha = \frac{K^*-\bar{K}}{\bar{k}-\bar{K}}$ into (33) yields

$$\begin{aligned} &(1-\alpha)\|y_{(\bar{K}+1):K^*}\|_2^2 \le \alpha \|\varepsilon_{(K^*+1):\bar{k}}\|_2^2 \\ \Rightarrow \quad &\|y_{(\bar{K}+1):K^*}\|_2^2 \le \frac{\alpha}{1-\alpha}\|\varepsilon_{(K^*+1):\bar{k}}\|_2^2 \\ &\qquad < \frac{\alpha}{1-\alpha}\lambda^2(\bar{k}-K^*) \\ &\qquad = \lambda^2(K^*-\bar{K}), \end{aligned}$$

where the last equality is by $\frac{\alpha}{1-\alpha}(\bar{k}-K^*) = K^*-\bar{K}$. Having established that (32) holds, we have $\forall k > K^*$ that

$$(34)\qquad \begin{aligned} \|y_{(\bar{K}+1):k}\|_2^2 &= \|y_{(\bar{K}+1):K^*}\|_2^2 + \|\varepsilon_{(K^*+1):k}\|_2^2 \\ &< \lambda^2(K^*-\bar{K}) + \lambda^2(k-K^*) \\ &= \lambda^2(k-\bar{K}) \end{aligned}$$

By (34) we have

$$(35)\qquad \|y_{(\bar{K}+1):k}\|_2^2 \le \lambda^2(k-\bar{K}) \quad \Leftrightarrow \quad f(k,\bar{K}) \le \lambda.$$

According to Algorithm 3, $\bar{K}$ is the last knot on the entire path graph and $\mathrm{supp}(\hat{\beta}^{\mathrm{LOG}}) \subseteq \mathrm{supp}(\beta^*)$. $\square$

LEMMA 8. *Under the assumptions of Proposition 2, the following holds on* $\mathcal{B}^c$: *For* $1 \le d \le d+h \le K^*$ *and* $\hat{\beta}_{d+h}^{\mathrm{LOG}} \neq 0$,

$$\delta \frac{|y_{d+h}|}{|y_d|} \le \frac{|\hat{\beta}_{d+h}^{\mathrm{LOG}}|}{|\hat{\beta}_d^{\mathrm{LOG}}|} \le \frac{|y_{d+h}|}{|y_d|}.$$

PROOF. For $1 \le d \le d+h \le K^*$ and $\hat{\beta}^{\rm LOG}_{d+h} \ne 0$, by Algorithm 3 we have

$$\frac{|\hat{\beta}^{\rm LOG}_{d+h}|}{|\hat{\beta}^{\rm LOG}_{d}|} = \frac{|y_{d+h}|}{|y_d|} \cdot \frac{1 - \frac{\lambda}{f(k^U(d+h), k^L(d+h))}}{1 - \frac{\lambda}{f(k^U(d), k^L(d))}}, \tag{36}$$

where $f(k, j) = \|y_{(j+1):k}\|_2/\sqrt{k-j}$ and $k^L(d)$ and $k^U(d)$ are two adjacent knots determined by Algorithm 3 such that $k^L(d) < d \le k^U(d)$ [and similarly $k^L(d+h) < d+h \le k^U(d+h)$]. For simplicity of notation, we denote $a := f(k^U(d+h), k^L(d+h))$ and $b := f(k^U(d), k^L(d))$.

By (36), we wish to show that

$$\delta \le \frac{1 - \lambda/a}{1 - \lambda/b} \le 1.$$

When $k^L(d) = k^L(d+h)$ and $k^U(d) = k^U(d+h)$, $a = b$ and thus this is immediate. It remains to consider the case when $k^L(d) < k^U(d) \le k^L(d+h) < k^U(d+h)$. By Lemma 9, we have that $b \ge a$, which gives the upper bound.

Some algebra shows that

$$\frac{1 - \lambda/a}{1 - \lambda/b} \ge \delta \quad \Leftrightarrow \quad \lambda \le \frac{1 - \delta}{1/a - \delta/b}. \tag{37}$$

We will show that the upper bound on $\lambda$ assumed in Proposition 2 ensures that the above inequality holds on $\mathcal{B}^c$.

For any $0 \le j < k \le K^*$,

$$\min_{i \in \{j+1, \dots, k\}} y_i^2 \le \|y_{(j+1):k}\|_2^2/(k-j) = f(k, j)^2$$

and thus on $\mathcal{B}^c$,

$$\begin{aligned} f(k, j) &\ge \min_{1 \le i \le K^*} |y_i| \\ &\ge 1 - \max_{1 \le i \le K^*} |\varepsilon_i| \quad \text{by the triangle inequality} \qquad (38) \\ &\ge 1 - \bar{\lambda} \quad \text{by definition of } \mathcal{B}^c. \end{aligned}$$

Since $\mathrm{supp}(\hat{\beta}^{\rm LOG}) \subseteq \mathrm{supp}(\beta^*)$ on $\mathcal{B}^c$ by Lemma 7 and $\hat{\beta}^{\rm LOG}_{d+h} \ne 0$ by assumption, we have $k^L(d+h) < k^U(d+h) \le K^*$. Taking $(k, j) = (k^U(d+h), k^L(d+h))$ in (38) yields

$$1 - \bar{\lambda} \le a \le \frac{1}{1/a - \delta/b}.$$

Thus, recalling the upper bound for $\lambda$ given in Proposition 2,

$$\lambda \le (1-\delta)(1-\bar{\lambda}) \le \frac{1-\delta}{1/a - \delta/b},$$

which by (37), establishes that

$$\frac{|\hat{\beta}^{\rm LOG}_{d+h}|}{|\hat{\beta}^{\rm LOG}_{d}|} \ge \frac{|y_{d+h}|}{|y_d|} \cdot \delta. \qquad \square$$

## APPENDIX C: PROOF THAT ALGORITHM 3 SOLVES $\mathrm{Prox}^{a(\mathcal{D})}_{\mathrm{LOG}}$ FOR A DIRECTED PATH GRAPH

Suppose $\mathcal{D}$ is a directed path graph with $D$ nodes as shown in Figure 4. Let $\hat{\beta} = \mathrm{Prox}^{a(\mathcal{D})}_{\mathrm{LOG}}(y; \lambda', w')$ and $\bar{\beta}$ denote the output from Algorithm 3 with inputs $\lambda'$ and $w'$. To prove $\bar{\beta} = \hat{\beta}$, we propose a $\{\bar{v}^{(\ell)}\}_{\ell=1}^D$ such that $\mathrm{supp}(\bar{v}^{(\ell)}) \subseteq s_{1:\ell}$ and $\bar{v}^{(\ell)} \in \mathbb{R}^p$ for $\ell = 1, \dots, D$. We then show that $\bar{\beta} = \sum_{\ell=1}^D \bar{v}^{(\ell)}$ and

$$\begin{cases} \bar{\beta}_{s_{1:\ell}} - y_{s_{1:\ell}} = -\dfrac{\lambda' w'_\ell \bar{v}^{(\ell)}}{\|\bar{v}^{(\ell)}\|_2} & \text{if } \bar{v}^{(\ell)} \ne 0, \\ \|\bar{\beta}_{s_{1:\ell}} - y_{s_{1:\ell}}\|_2 \le \lambda' w'_\ell & \text{if } \bar{v}^{(\ell)} = 0. \end{cases} \tag{39}$$

By the optimality condition stated in Lemma 11 of Obozinski, Jacob and Vert (2011), this establishes that $\bar{\beta} = \hat{\beta}$. Let $0 = k_0 < k_1 < \cdots < k_m \le D$ be the sequence of knots determined by Algorithm 3 such that $k_i$ maximizes $f(\cdot, k_{i-1})$ and $f(k_i, k_{i-1}) > \lambda'$ for $i = 1, \dots, m$.

If $m = 0$, that is, $k_0 = 0$ is the only knot, we have $\bar{\beta} = 0$. Consider $\bar{v}^{(\ell)} = 0$ for $\ell = 1, \dots, D$, which satisfy $\bar{\beta} = \sum_{\ell=1}^D \bar{v}^{(\ell)}$. Moreover, we get $\|y_{s_{1:\ell}}\|_2/w'_\ell \le \lambda'$ for $\ell = 1, \dots, D$ directly from the algorithm. By Lemma 11 of Obozinski, Jacob and Vert (2011), $\bar{\beta} = \hat{\beta}$.

Now consider $m \ge 1$. We first prove an inequality in $f(j, k)$ in Lemma 9 when $(k, j)$ are two nearest knots.

LEMMA 9. *Let $0 = k_0 < k_1 < \cdots < k_m \le D$ be the sequence of knots. We have the following inequality*:

$$f(k_{j-1}, k_{j-2}) \ge f(k_j, k_{j-1}) \quad \textit{for } j = 2, \dots, m.$$

PROOF. Applying Algorithm 3 yields that for $j = 2, \dots, m$,

$$\begin{aligned} & f(k_{j-1}, k_{j-2}) \ge f(k_j, k_{j-2}) \\ \Rightarrow\quad & \frac{\|y_{s_{(k_{j-2}+1):k_{j-1}}}\|_2}{\sqrt{w'^2_{k_{j-1}} - w'^2_{k_{j-2}}}} \ge \frac{\|y_{s_{(k_{j-2}+1):k_j}}\|_2}{\sqrt{w'^2_{k_j} - w'^2_{k_{j-2}}}} \\ \Rightarrow\quad & \frac{w'^2_{k_{j-1}} - w'^2_{k_{j-2}}}{\|y_{s_{(k_{j-2}+1):k_{j-1}}}\|_2^2} \le \frac{w'^2_{k_j} - w'^2_{k_{j-2}}}{\|y_{s_{(k_{j-2}+1):k_j}}\|_2^2} \\ \Rightarrow\quad & \frac{w'^2_{k_{j-1}} - w'^2_{k_{j-2}}}{\|y_{s_{(k_{j-2}+1):k_{j-1}}}\|_2^2} - \frac{w'^2_{k_{j-1}} - w'^2_{k_{j-2}}}{\|y_{s_{(k_{j-2}+1):k_j}}\|_2^2} \\ & \qquad \le \frac{w'^2_{k_j} - w'^2_{k_{j-2}}}{\|y_{s_{(k_{j-2}+1):k_j}}\|_2^2} - \frac{w'^2_{k_{j-1}} - w'^2_{k_{j-2}}}{\|y_{s_{(k_{j-2}+1):k_j}}\|_2^2} \end{aligned}$$

$$\Rightarrow \quad \frac{(w'^2_{k_{j-1}} - w'^2_{k_{j-2}})\|y_{s_{(k_{j-1}+1):k_j}}\|_2^2}{\|y_{s_{(k_{j-2}+1):k_{j-1}}}\|_2^2 \|y_{s_{(k_{j-2}+1):k_j}}\|_2^2} \le \frac{w'^2_{k_j} - w'^2_{k_{j-1}}}{\|y_{s_{(k_{j-2}+1):k_j}}\|_2^2}$$

$$\Rightarrow \quad \frac{w'^2_{k_{j-1}} - w'^2_{k_{j-2}}}{\|y_{s_{(k_{j-2}+1):k_{j-1}}}\|_2^2} \le \frac{w'^2_{k_j} - w'^2_{k_{j-1}}}{\|y_{s_{(k_{j-1}+1):k_j}}\|_2^2}$$

$$\Rightarrow \quad \frac{\sqrt{w'^2_{k_{j-1}} - w'^2_{k_{j-2}}}}{\|y_{s_{(k_{j-2}+1):k_{j-1}}}\|_2} \le \frac{\sqrt{w'^2_{k_j} - w'^2_{k_{j-1}}}}{\|y_{s_{(k_{j-1}+1):k_j}}\|_2}$$

$$\Rightarrow \quad \frac{1}{f(k_{j-1}, k_{j-2})} \le \frac{1}{f(k_j, k_{j-1})}$$

$$\Rightarrow \quad f(k_{j-1}, k_{j-2}) \ge f(k_j, k_{j-1}). \qquad \square$$

For notational simplicity, we let $a_j = f(k_j, k_{j-1})$ for $j = 1, \dots, m$, and let

$$A_j = \sum_{i=1}^{j} \frac{y_{s_{(k_{i-1}+1):k_i}}}{a_i}.$$

We observe that

$$\|A_j\|_2^2 = \sum_{i=1}^{j} \frac{\|y_{s_{(k_{i-1}+1):k_i}}\|_2^2}{a_i^2} = \sum_{i=1}^{j} (w'^2_{k_i} - w'^2_{k_{i-1}}) = w'^2_{k_j}. \tag{40}$$

Now consider the following $\{\bar{v}^{(\ell)}\}_{\ell=1}^D$ such that $\mathrm{supp}(\bar{v}^{(\ell)}) \subseteq s_{1:\ell}$ and $\bar{v}^{(\ell)} \in \mathbb{R}^p$ $\forall \ell$.

- For $\ell \notin \{k_1, \dots, k_m\}$,
$$\bar{v}^{(\ell)} = 0.$$
- For $\ell = k_j$ for $j = 1, \dots, m-1$,
$$\begin{aligned}\bar{v}^{(k_j)} &= S_G(a_j A_j, w'_{k_j} a_{j+1}) \\ &= A_j \cdot a_j \cdot \left(1 - \frac{w'_{k_j} a_{j+1}}{a_j \|A_j\|_2}\right)_+ \\ &= A_j \cdot (a_j - a_{j+1})\end{aligned}$$
by (40) and $a_j \ge a_{j+1}$ from Lemma 9.
- For $\ell = k_m$,
$$\begin{aligned}\bar{v}^{(k_m)} &= S_G(a_m A_m, \lambda' w'_{k_m}) \\ &= a_m A_m \cdot \left(1 - \frac{\lambda' w'_{k_m}}{a_m \|A_m\|_2}\right)_+ \\ &= A_m \cdot (a_m - \lambda')\end{aligned}$$
by $a_m > \lambda'$ from Algorithm 3.

Because of the very definition of $\bar{\beta}$ in Algorithm 3, we can express $\bar{\beta}$ in the following form:

- For $1 \le i \le m$,
$$\bar{\beta}_{s_{(k_{i-1}+1):k_i}} = S_G\left(y_{s_{(k_{i-1}+1):k_i}}, \lambda' \sqrt{w'^2_{k_i} - w'^2_{k_{i-1}}}\right).$$
- If $k_m < D$, $\bar{\beta}_{s_{(k_m+1):D}} = 0$.

We show that $\bar{\beta} = \sum_{\ell=1}^D \bar{v}^{(\ell)}$ through steps (a), (b) and (c) below.

(a) For $i = 1, \dots, m-1$,

$$\begin{aligned}\sum_{\ell=1}^{D} \bar{v}^{(\ell)}_{s_{(k_{i-1}+1):k_i}} &= \sum_{j=i}^{m} \bar{v}^{(k_j)}_{s_{(k_{i-1}+1):k_i}} \\ &= \sum_{j=i}^{m-1} \bar{v}^{(k_j)}_{s_{(k_{i-1}+1):k_i}} + \bar{v}^{(k_m)}_{s_{(k_{i-1}+1):k_i}} \\ &= \frac{y_{s_{(k_{i-1}+1):k_i}}}{a_i} \sum_{j=i}^{m-1} (a_j - a_{j+1}) \\ &\quad + \frac{y_{s_{(k_{i-1}+1):k_i}}}{a_i} (a_m - \lambda') \\ &= \frac{y_{s_{(k_{i-1}+1):k_i}}}{a_i} (a_i - \lambda') \\ &= y_{s_{(k_{i-1}+1):k_i}} \left(1 - \frac{\lambda'}{a_i}\right) \\ &= S_G\left(y_{s_{(k_{i-1}+1):k_i}}, \lambda' \sqrt{w'^2_{k_i} - w'^2_{k_{i-1}}}\right) \\ &= \bar{\beta}_{s_{(k_{i-1}+1):k_i}}.\end{aligned}$$

(b) For $i = m$,

$$\begin{aligned}\sum_{\ell=1}^{D} \bar{v}^{(\ell)}_{s_{(k_{i-1}+1):k_i}} &= \bar{v}^{(k_m)}_{s_{(k_{m-1}+1):k_m}} \\ &= \frac{y_{s_{(k_{m-1}+1):k_m}}}{a_m} (a_m - \lambda') \\ &= y_{s_{(k_{m-1}+1):k_m}} \left(1 - \frac{\lambda'}{a_m}\right) \\ &= S_G\left(y_{s_{(k_{m-1}+1):k_m}}, \lambda' \sqrt{w'^2_{k_m} - w'^2_{k_{m-1}}}\right) \\ &= \bar{\beta}_{s_{(k_{m-1}+1):k_m}}.\end{aligned}$$

(c) If $k_m < D$, $\sum_{\ell=1}^D \bar{v}^{(\ell)}_{s_{(k_m+1):D}} = 0 = \bar{\beta}_{s_{(k_m+1):D}}$.

Combining (a), (b) and (c) we have established $\bar{\beta} = \sum_{\ell=1}^D \bar{v}^{(\ell)}$. We next show (39) is true through steps (a′) and (b′) below.

(a′) By definition, $\bar{v}^{(\ell)} \neq 0$ if and only if $\ell \in \{k_1, \dots, k_m\}$. For $\ell = k_i \in \{k_1, \dots, k_m\}$, we have

$$\begin{aligned}
&\bar{\beta}_{s_{1:k_i}} - y_{s_{1:k_i}} \\
&\quad = \sum_{j=1}^{i} S_G\Big(y_{s_{(k_{j-1}+1):k_j}}, \lambda' \sqrt{w_{k_j}'^2 - w_{k_{j-1}}'^2}\Big) \\
&\qquad - y_{s_{1:k_i}} \\
&\quad = \sum_{j=1}^{i} y_{s_{(k_{j-1}+1):k_j}} (1 - \lambda' a_j^{-1}) - y_{s_{1:k_i}} \\
&\quad = \sum_{j=1}^{i} - \frac{\lambda' y_{s_{(k_{j-1}+1):k_j}}}{a_j} = -\lambda' A_i.
\end{aligned}$$

By the definition of $\{\bar{v}^{(\ell)}\}_{\ell=1}^{D}$, we have

$$-\frac{\lambda' w_{k_i}' \bar{v}^{(k_i)}}{\|\bar{v}^{(k_i)}\|_2} = -\frac{\lambda' w_{k_i}' A_i}{\|A_i\|_2} = -\lambda' A_i.$$

Thus, $\bar{\beta}_{s_{1:\ell}} - y_{s_{1:\ell}} = -\frac{\lambda' w_\ell' \bar{v}^{(\ell)}}{\|\bar{v}^{(\ell)}\|_2}$ if $\bar{v}^{(\ell)} \neq 0$.

(b′) By definition, $\bar{v}^{(\ell)} = 0$ if and only if $\ell \notin \{k_1, \dots, k_m\}$. We discuss $\ell$ in the following three cases.

(i) If $k_{i-1} < \ell < k_i$ for some $i = 2, \dots, m$, by Algorithm 3 we have

$$\bar{\beta}_{s_{1:\ell}} - y_{s_{1:\ell}} = -\lambda' A_{i-1} - \frac{\lambda' y_{s_{(k_{i-1}+1):\ell}}}{a_i}.$$

Taking $\ell_2$-norm on both sides yields

$$\|\bar{\beta}_{s_{1:\ell}} - y_{s_{1:\ell}}\|_2 = \lambda' \sqrt{w_{k_{i-1}}'^2 + \frac{(w_{k_i}'^2 - w_{k_{i-1}}'^2)\|y_{s_{(k_{i-1}+1):\ell}}\|_2^2}{\|y_{s_{(k_{i-1}+1):k_i}}\|_2^2}}. \tag{41}$$

By the Algorithm 3, we know

$$k_i = \operatorname*{arg\,max}_{i' \in \{k_{i-1}+1, \dots, D\}} f(i', k_{i-1}),$$

so that $a_i \geq f(\ell, k_{i-1})$ which leads to

$$\begin{aligned}
&\frac{\|y_{s_{(k_{i-1}+1):k_i}}\|_2^2}{(w_{k_i}'^2 - w_{k_{i-1}}'^2)} \geq \frac{\|y_{s_{(k_{i-1}+1):\ell}}\|_2^2}{(w_\ell'^2 - w_{k_{i-1}}'^2)} \\
&\quad \Rightarrow \quad \frac{(w_{k_i}'^2 - w_{k_{i-1}}'^2)\|y_{s_{(k_{i-1}+1):\ell}}\|_2^2}{\|y_{s_{(k_{i-1}+1):k_i}}\|_2^2} \\
&\qquad \leq w_\ell'^2 - w_{k_{i-1}}'^2.
\end{aligned} \tag{42}$$

Combining (41) and (42) yields

$$\begin{aligned}
\|\bar{\beta}_{s_{1:\ell}} - y_{s_{1:\ell}}\|_2 &\leq \lambda' \sqrt{w_{k_{i-1}}'^2 + w_\ell'^2 - w_{k_{i-1}}'^2} \\
&= \lambda' w_\ell'.
\end{aligned}$$

(ii) If $\ell < k_1$, $\bar{\beta}_{s_{1:\ell}} - y_{s_{1:\ell}} = -\lambda' y_{s_{1:\ell}}/a_1$. Since $k_1 = \arg\max_{i' \in \{1, \dots, D\}} f(i', 0)$, we have $a_1 \geq f(\ell, 0)$ which leads to

$$\begin{aligned}
&\frac{\|y_{s_{1:k_1}}\|_2^2}{w_{k_1}'^2} \geq \frac{\|y_{s_{1:\ell}}\|_2'^2}{w_\ell'^2} \\
&\quad \Rightarrow \quad \frac{w_{k_1}'^2 \|y_{s_{1:\ell}}\|_2^2}{\|y_{s_{1:k_1}}\|_2^2} \leq w_\ell'^2.
\end{aligned} \tag{43}$$

By (43) we get

$$\|\bar{\beta}_{s_{1:\ell}} - y_{s_{1:\ell}}\|_2 = \sqrt{\frac{\lambda'^2 w_{k_1}'^2 \|y_{s_{1:\ell}}\|_2^2}{\|y_{s_{1:k_1}}\|_2^2}} \leq \lambda' w_\ell'.$$

(iii) If $\ell > k_m$ (provided $k_m < D$),

$$\bar{\beta}_{s_{1:\ell}} - y_{s_{1:\ell}} = -\lambda' A_m - y_{s_{(k_m+1):\ell}}.$$

Since $k_m$ is the last knot, we know that

$$\max_{i' \in \{k_m+1, \dots, D\}} f(i', k_m) \leq \lambda'.$$

Thus, $f(\ell, k_m) \leq \lambda'$ which leads to

$$\|y_{s_{(k_m+1):\ell}}\|_2^2 \leq \lambda'^2 (w_\ell'^2 - w_{k_m}'^2).$$

Thus,

$$\begin{aligned}
\|\bar{\beta}_{s_{1:\ell}} - y_{s_{1:\ell}}\|_2 &= \sqrt{\lambda'^2 \|A_m\|_2^2 + \|y_{s_{(k_m+1):\ell}}\|_2^2} \\
&= \sqrt{\lambda'^2 w_{k_m}'^2 + \|y_{s_{(k_m+1):\ell}}\|_2^2} \\
&\leq \sqrt{\lambda'^2 w_{k_m}'^2 + \lambda'^2 (w_\ell'^2 - w_{k_m}'^2)} \\
&= \lambda' w_\ell'.
\end{aligned}$$

Combining (a′) and (b′) we prove (39) holds. Since the second optimality condition in Lemma 11 of Obozinski, Jacob and Vert (2011) is satisfied, we have $\bar{\beta} = \hat{\beta}$.

## APPENDIX D: COMPUTATIONAL COMPLEXITY OF ALGORITHM 3

Let $z_i = \|y_{s_i}\|_2^2$ for $i = 1, \dots, D$. We begin by computing all the $z_i$, which takes $O(p)$ operations. To compute the $i$th knot requires computing $f(j, k_{i-1})$ for $j = k_{i-1} + 1, \dots, D$.

To compute $f(k+1, k)^2 = z_{k+1}/(w_{k+1}^2 - w_k^2)$ requires constant time; also, once $f(j, k)$ has been computed, we can get $f(j+1, k)$ in constant time since

$$f(j+1, k)^2 = \frac{(w_j^2 - w_k^2) f(j, k)^2 + z_{j+1}}{(w_{j+1}^2 - w_k^2)}.$$

Thus computing all the $f(\cdot, k_{i-1})$'s requires $O(D - k_{i-1})$ operations. Finding the maximizer in line 5 takes an additional $O(D - k_{i-1})$ operations. Thus, in total finding all knots requires on the order of

$$p + \sum_{i=1}^{m}(D - k_{i-1})$$

operations. Once the knots have been found, the group-wise soft-thresholding steps require only an additional $O(p)$ work. Therefore, the algorithm requires $O(p + mD)$ operations. Since the number of knots is not known *a priori*, the worst case is $O(p + D^2)$.

## APPENDIX E: COMPUTATIONAL COMPLEXITY OF GL FOR A DIRECTED PATH GRAPH

### E.1 GL Proximal Operator

By Jenatton et al.'s (2011) result, Algorithm 1 will converge in a single pass when $\mathcal{D}$ is a directed path graph if we cycle through the groups $g_i = s_{(D+1-i):D}$ from smallest to largest. The algorithm can be stated simply as follows: Initialize $\beta^0 = y$ and then for $i = 1, \dots, D$, set

$$\beta_{g_i}^{i} \leftarrow \left(1 - \frac{\lambda w_i}{\|\beta_{g_i}^{i-1}\|_2}\right)_+ \beta_{g_i}^{i-1},$$

and output $\beta^D$ as the solution. As in Appendix D, we begin by computing $z_i = \|y_{s_i}\|_2^2$ for $i = 1, \dots, D$, which can be done in $O(p)$ operations. Define $a_i = \|\beta_{g_i}^{i-1}\|_2^2$ and observe that $a_1 = z_1$ and that, for $i \geq 1$,

$$a_{i+1} = z_{i+1} + \|\beta_{g_i}^{i}\|_2^2 = z_{i+1} + (a_i^{1/2} - \lambda w_i)_+^2.$$

Thus, we can compute $a_1, \dots, a_D$ in $O(D)$ operations. For $\ell = 1, \dots, D$, we form $b_\ell = \prod_{i=\ell}^{D}(1 - \frac{\lambda w_i}{\sqrt{a_i}})_+$ [which can be done in $O(D)$ operations] and observe that

$$\beta_{s_\ell}^{D} = b_\ell y_{s_\ell}.$$

This final scaling of the elements of $y$ takes $O(p)$. Thus, computing the GL proximal operator can be done in $O(p + D)$ operations.

### E.2 Modified GL Proximal Operator

When we introduced $\Omega_{\text{mGL}}^{d(\mathcal{D})}$ in (16) of Section 3, we defined the penalty in the one parameter per node case. Following Bien, Bunea and Xiao (2016), we now generalize the definition to the situation of multiple parameters per node in a directed path graph $\mathcal{D}$. For $\ell = 1, \dots, D$, we let $g_\ell = s_{\ell:D}$. Let $w_{\ell,m} = \frac{\sqrt{|s_\ell|}}{m-\ell+1}$

---

**Algorithm 5** Solve proximal operator of modified GL in (44)

$\beta^{D+1} \leftarrow y$
**for** $i = D, \dots, 1$ **do**
  Solve $\lambda^2 = \sum_{m=i}^{D} \frac{w_{i,m}^2}{(w_{i,m}^2 + \hat{\nu}^{(i)})^2}\|\beta_{s_m}^{i+1}\|_2^2$ for $\hat{\nu}^{(i)}$
  **for** $m = 1, \dots, D$ **do**
    $\beta_{s_m}^{i} \leftarrow \frac{[\hat{\nu}^{(i)}]_+}{w_{i,m}^2 + [\hat{\nu}^{(i)}]_+}\beta_{s_m}^{i+1}$
  **end for**
**end for**
**Output:** $\beta^1$

---

where $1 \leq \ell \leq m \leq D$ be the weight applied to $s_m$ in $g_\ell$. The modified GL penalty under a path graph can be written as

$$\Omega_{\text{mGL}}^{d(\mathcal{D})}(\beta; \{w_{\ell,m}\}) = \sum_{\ell=1}^{D} \sqrt{\sum_{m=\ell}^{D} w_{\ell,m}^2 \|\beta_{s_m}\|_2^2}, \tag{44}$$

By Jenatton et al.'s (2011) result, a single pass of BCD from $g_D$ to $g_1$ will solve the dual problem. Bien, Bunea and Xiao (2016) proves the modified version of BCD in the context of covariance estimation, which itself is a special case of directed path graphs. By Theorem 2 of Bien, Bunea and Xiao (2016), we have the algorithm stated in Algorithm 5.

We can define $t \in \mathbb{R}^p$ such that for $m = 1, \dots, D$,

$$(t_{s_m})_j = \begin{cases} \sum_{i=1}^{m} \frac{[\hat{\nu}^{(i)}]_+}{w_{i,m}^2 + [\hat{\nu}^{(i)}]_+} & \text{if } j \in s_m, \\ 0 & \text{otherwise.} \end{cases}$$

The solution $\hat{\beta}$ can be written as $\hat{\beta} = t * y$ where $*$ denotes elementwise multiplication. Provided all the $\{\hat{\nu}^{(i)}\}_{i=1,\dots D}$ have been found, computing $t$ requires $O(\sum_{m=1}^{D} m) = O(D^2)$ operations. Performing elementwise multiplication to get $\hat{\beta}$ can be done in $O(p)$ operations.

To find a root $\{\hat{\nu}^{(i)}\}_{i=1,\dots D}$, Bien, Bunea and Xiao (2016) shows that $\hat{\nu}^{(i)} \leq 0$ when $\lambda^2 \geq \sum_{m=i}^{D} \|\beta_{s_m}^{i+1}\|_2^2 / w_{i,m}^2$. In that case, $\beta_{g_i}^{i} = 0$. If parameters corresponding to $\{g_D, \dots, g_{\hat{K}+1}\}$ are zeroed out, only the last $\hat{K}$ roots need to be numerically computed. We start by computing $z_i = \|y_{s_i}\|_2^2$ for $i = 1, \dots, D$, which can be done in $O(p)$ operations. Then do the following two steps:

1. Compute $z_i/|s_i|$ for $i = D, \dots, 1$. Let $i = \hat{K}$ be the first time $\lambda^2 < z_i/|s_i|$. The amount of operations is $O(D)$. At the end of this part, we have $\beta_{g_{\hat{K}+1}}^{\hat{K}+1} = 0$ if $\hat{K} < D$.

2. For $i \in \{\hat{K}, \dots, 1\}$, we need to find $\nu$ such that

$$f(\nu) = 1 - \frac{\lambda}{\sqrt{\sum_{m=i}^{D} \frac{w_{i,m}^2 \|\beta_{s_m}^{i+1}\|_2^2}{(w_{i,m}^2+\nu)^2}}} = 1 - \frac{\lambda}{\sqrt{\sum_{m=i}^{\hat{K}} \frac{w_{i,m}^2 \|\beta_{s_m}^{i+1}\|_2^2}{(w_{i,m}^2+\nu)^2}}} = 0,$$

which can be solved using Newton's method. At each iteration of Newton's method, we need to compute

$$\frac{f(\nu)}{f'(\nu)} = \left(\sum_{m=i}^{\hat{K}} \frac{w_{i,m}^2 \|\beta_{s_m}^{i+1}\|_2^2}{(w_{i,m}^2+\nu)^2} - \lambda^{-1}\left(\sum_{m=i}^{\hat{K}} \frac{w_{i,m}^2 \|\beta_{s_m}^{i+1}\|_2^2}{(w_{i,m}^2+\nu)^2}\right)^{1.5}\right) \Big/ \sum_{m=i}^{\hat{K}} \frac{w_{i,m}^2 \|\beta_{s_m}^{i+1}\|_2^2}{(w_{i,m}^2+\nu)^3}.$$

Evaluating $\|\beta_{s_m}^{i+1}\|_2^2$ can be done efficiently. For $i = \hat{K}, \dots, 1$ and $m = i, \dots, \hat{K}$, define $a^{(i,m)} = \|\beta_{s_m}^{i+1}\|_2^2$. It is obvious that $a^{(i,i)} = \|y_{s_i}\|_2^2 = z_i$ for $i = \hat{K}, \dots, 1$. For $m \ge i$, we have

$$a^{(i-1,m)} = \|\beta_{s_m}^{i}\|_2^2 = \left(\frac{[\hat{\nu}^{(i)}]_+}{w_{i,m}^2 + [\hat{\nu}^{(i)}]_+}\right)^2 a^{(i,m)}.$$

Applying this update, we can compute all $\{a^{(i,m)}\}$ with $i \le m$ in a total of $O(\sum_{m=1}^{\hat{K}} m) = O(\hat{K}^2)$ operations. At a fixed $i = \hat{K}, \dots, 1$, provided all the needed $\{a^{(i,m)}\}$ are computed already, evaluating $f(\nu)/f'(\nu)$ requires $O(\hat{K} - i)$ per $\nu$ value. Newton's method is known for its quadratic convergence rate once the estimate gets "near" a root (Proposition 1.4.1 of Bertsekas, 1999). Therefore, the number of significant digits double with each iteration when the estimate gets close to the root. For $n$-digit precision, Newton's method needs $O(\log(n) \cdot (\hat{K} - i))$ operations if the initial point is good. Therefore, the total amount of computations for Step 2 is

$$O\left(\hat{K}^2 + \log(n)\sum_{i=1}^{\hat{K}}(\hat{K} - i)\right) = O(\log(n)\hat{K}^2) = O(D^2 \log(n)).$$

Combing the above derivation, the proximal operator of modified GL can be computed in $O(p + D^2 \log(n))$ operations, where $n$ is the pre-determined number of digits of precision for Newton's method.

## APPENDIX F: PROOF OF LEMMA 4

Recalling that $\mathcal{G}_1, \dots, \mathcal{G}_L$ is a partition of $a(\mathcal{D})$, we can write Problem (8) as the following:

$$\min_{\beta \in \mathbb{R}^p} \{F(\beta) + \lambda \Omega_{\text{LOG}}^{a(\mathcal{D})}(\beta; w)\}$$

$$\Leftrightarrow \min_{\{v^{(g)} \in \mathbb{R}^p\}_{g \in a(\mathcal{D})}} \left\{ F\left(\sum_{\ell=1}^{L} \sum_{g \in \mathcal{G}_\ell} v^{(g)}\right) + \lambda \sum_{\ell=1}^{L} \sum_{g \in \mathcal{G}_\ell} w_g \|v^{(g)}\|_2 \quad \text{s.t. } v_{g^c}^{(g)} = 0\ \forall g \in a(\mathcal{D}) \right\} \tag{45}$$

$$\Leftrightarrow \min_{\{v^{(g)} \in \mathbb{R}^p\}_{g \in a(\mathcal{D})}} \left\{ F\left(\sum_{\ell=1}^{L} \beta^{(\ell)}\right) + \lambda \sum_{\ell=1}^{L} \sum_{g \in \mathcal{G}_\ell} w_g \|v^{(g)}\|_2 \quad \text{s.t. } v_{g^c}^{(g)} = 0\ \forall g \in a(\mathcal{D}), \beta^{(\ell)} = \sum_{g \in \mathcal{G}_\ell} v^{(g)} \right\}.$$

Finally, by definition of the LOG penalty, we can write (45) as

$$\min_{\{\beta^{(\ell)} \in \mathbb{R}^p\}_{\ell=1}^{L}} \left\{ F\left(\sum_{\ell=1}^{L} \beta^{(\ell)}\right) + \lambda \sum_{\ell=1}^{L} \Omega_{\text{LOG}}^{\mathcal{G}_\ell}(\beta^{(\ell)}; w_{\mathcal{P}_\ell}) \quad \text{s.t. } \text{supp}(\beta^{(\ell)}) \subset \bigcup_{g \in \mathcal{G}_\ell} g \right\},$$

where $w_{\mathcal{P}_\ell} = \{w_g : g \in \mathcal{G}'_\ell\}$.

## APPENDIX G: SIMPLE ALGORITHM FOR PATH DECOMPOSITION OF DAG

Algorithm 6 presents a simple greedy algorithm for decomposing $\mathcal{D}$ into paths.

**Algorithm 6** Path decomposition of a DAG $\mathcal{D}$

**Input:** $\mathcal{D}$

$\mathcal{M} \leftarrow \varnothing$ and $L \leftarrow 1$
Form set of "root nodes" $R = \{s_i : \text{ancestors}(\mathcal{D}; s_i) = \{s_i\}\}$.
**for** $s_i \in R$ **do**
  **while** descendants$(\mathcal{D}; s_i) \nsubseteq \mathcal{M}$ **do**
    Choose the path $\mathcal{P}$ from $s_i$ for which $|\mathcal{P} \setminus \mathcal{M}|$ is largest.
    Define $\mathcal{P}_\ell \leftarrow \mathcal{P} \setminus \mathcal{M}$
    $\mathcal{M} \leftarrow \mathcal{M} \cup \mathcal{P}_\ell$.
    $L \leftarrow L + 1$
  **end while**
**end for**

**Output:** $\mathcal{P}_1, \ldots, \mathcal{P}_L$.

## APPENDIX H: PROOF OF LEMMA 5

By Lemma 4, Problem (8) with

$$F(\beta) = \frac{1}{2}\|y - X\beta\|_2^2$$

can be written in terms of $\{\beta^{(\ell)}\}_{\ell=1}^L$ subject to $\beta = \sum_{\ell=1}^L \beta^{(\ell)}$:

$$\begin{aligned}&\min_{\{\beta^{(\ell)}\in\mathbb{R}^p\}_{\ell=1}^L} \frac{1}{2}\left\| y - \mathbf{X}\sum_{\ell=1}^L \beta^{(\ell)}\right\|_2^2 \\ &\qquad + \lambda \sum_{\ell=1}^L \Omega_{\text{LOG}}^{\mathcal{G}_\ell}(\beta^{(\ell)}; w_{\mathcal{P}_\ell}) \\ &\qquad \text{s.t. } \operatorname{supp}(\beta^{(\ell)}) \subseteq g^{(\ell)}\ \forall \ell = 1, \ldots, L.\end{aligned} \tag{46}$$

Then (19) follows by substituting $\{\beta^{(\ell)}\}$ with $\{\gamma^{(\ell)}\}$ in the squared loss of (46). The augmented Lagrangian subject to $\operatorname{supp}(\beta^{(\ell)}) \subseteq g^{(\ell)}$ and $\operatorname{supp}(\gamma^{(\ell)}) \subseteq g^{(\ell)}$ $\forall\ell$ is

$$\begin{aligned}&L(\{\beta^{(\ell)}\}, \{\gamma^{(\ell)}\}, \{u^{(\ell)}\}) \\ &\quad = \frac{1}{2}\left\| y - \mathbf{X}\sum_{\ell=1}^L \gamma^{(\ell)}\right\|_2^2 + \lambda \sum_{\ell=1}^L \Omega_{\text{LOG}}^{\mathcal{G}_\ell}(\beta^{(\ell)}; w_{\mathcal{P}_\ell}) \\ &\qquad + \left\langle \begin{pmatrix} u^{(1)} \\ \vdots \\ u^{(L)}\end{pmatrix}, \begin{pmatrix} \beta^{(1)} - \gamma^{(1)} \\ \vdots \\ \beta^{(L)} - \gamma^{(L)}\end{pmatrix}\right\rangle \\ &\qquad + \frac{\rho}{2}\left\| \begin{pmatrix} \beta^{(1)} - \gamma^{(1)} \\ \vdots \\ \beta^{(L)} - \gamma^{(L)}\end{pmatrix}\right\|_2^2 \\ &\quad = \frac{1}{2}\left\| y - \mathbf{X}\sum_{\ell=1}^L \gamma^{(\ell)}\right\|_2^2 + \lambda \sum_{\ell=1}^L \Omega_{\text{LOG}}^{\mathcal{G}_\ell}(\beta^{(\ell)}; w_{\mathcal{P}_\ell}) \\ &\qquad + \frac{\rho}{2}\sum_{\ell=1}^L \left\| \beta^{(\ell)} - \gamma^{(\ell)} + \frac{1}{\rho}u^{(\ell)}\right\|_2^2 \\ &\qquad - \frac{1}{2\rho}\sum_{\ell=1}^L \|u^{(\ell)}\|_2^2.\end{aligned}$$

Alternating Direction Method of Multipliers (ADMM) iteratively updates $\{\gamma^{(\ell)}\}$ and $\{\beta^{(\ell)}\}$ by optimizing the corresponding part in the augmented Lagrangian.

*Step 1*: Optimize over $\{\gamma^{(\ell)}\}$. For $\ell = 1, \ldots, L$,

$$\begin{aligned}\hat{\gamma}^{(\ell)} = \underset{\gamma^{(\ell)}\in\mathbb{R}^p}{\arg\min}\ &\frac{1}{2}\left\| y - \mathbf{X}\sum_{\ell'=1}^L \gamma^{(\ell')}\right\|_2^2 \\ &+ \frac{\rho}{2}\left\| \hat{\beta}^{(\ell)} - \gamma^{(\ell)} + \frac{1}{\rho}\hat{u}^{(\ell)}\right\|_2^2 \\ &\text{s.t. } \operatorname{supp}(\gamma^{(\ell)}) \subseteq g^{(\ell)}.\end{aligned}$$

Solving the gradient with respect to $\gamma_{|g^{(\ell)}}^{(\ell)}$ equal to zero yields

$$\begin{aligned}&\mathbf{X}_{|g^{(\ell)}}^T\Big(\mathbf{X}\sum_{\ell'}\gamma^{(\ell')} - y\Big) \\ &\qquad + \rho\Big(\gamma_{|g^{(\ell)}}^{(\ell)} - \hat{\beta}_{|g^{(\ell)}}^{(\ell)} - \frac{1}{\rho}\hat{u}_{|g^{(\ell)}}^{(\ell)}\Big) = 0.\end{aligned}$$

It follows that

$$\begin{aligned}\gamma_{|g^{(\ell)}}^{(\ell)} &= \hat{\beta}_{|g^{(\ell)}}^{(\ell)} + \frac{1}{\rho}\hat{u}_{|g^{(\ell)}}^{(\ell)} \\ &\qquad + \frac{1}{\rho}\mathbf{X}_{|g^{(\ell)}}^T\Big(y - \mathbf{X}\sum_{\ell'}\gamma^{(\ell')}\Big) \\ &= \hat{\beta}_{|g^{(\ell)}}^{(\ell)} + \frac{1}{\rho}\hat{u}_{|g^{(\ell)}}^{(\ell)} \\ &\qquad + \frac{1}{\rho}\mathbf{X}_{|g^{(\ell)}}^T\Big(y - \sum_{\ell'}\mathbf{X}_{|g^{(\ell')}}\gamma_{|g^{(\ell')}}^{(\ell')}\Big).\end{aligned} \tag{47}$$

Left-multiplying both sides of (47) by $\mathbf{X}_{|g^{(\ell)}}$ yields

$$\begin{aligned}\mathbf{X}_{|g^{(\ell)}}\gamma_{|g^{(\ell)}}^{(\ell)} &= \mathbf{X}_{|g^{(\ell)}}\Big(\hat{\beta}_{|g^{(\ell)}}^{(\ell)} + \frac{1}{\rho}\hat{u}_{|g^{(\ell)}}^{(\ell)}\Big) \\ &\qquad + \frac{1}{\rho}\mathbf{X}_{|g^{(\ell)}}\mathbf{X}_{|g^{(\ell)}}^T\Big(y - \sum_{\ell'}\mathbf{X}_{|g^{(\ell')}}\gamma_{|g^{(\ell')}}^{(\ell')}\Big).\end{aligned} \tag{48}$$

Summing up (48) over all $\ell$'s yields

$$\begin{aligned}
&\sum_\ell \mathbf{X}_{|g^{(\ell)}}\gamma^{(\ell)}_{|g^{(\ell)}} \\
&\quad = \sum_\ell \Big[\mathbf{X}_{|g^{(\ell)}}\Big(\hat{\beta}^{(\ell)}_{|g^{(\ell)}} + \frac{1}{\rho}\hat{u}^{(\ell)}_{|g^{(\ell)}}\Big) \\
&\qquad + \frac{1}{\rho}\mathbf{X}_{|g^{(\ell)}}\mathbf{X}^T_{|g^{(\ell)}} y\Big] \\
&\qquad - \frac{1}{\rho}\sum_\ell \mathbf{X}_{|g^{(\ell)}}\mathbf{X}^T_{|g^{(\ell)}}\sum_{\ell'}\mathbf{X}_{|g^{(\ell')}}\gamma^{(\ell')}_{|g^{(\ell')}} \\
&\quad \Rightarrow \quad \Big(I + \frac{1}{\rho}\sum_\ell \mathbf{X}_{|g^{(\ell)}}\mathbf{X}^T_{|g^{(\ell)}}\Big)\sum_\ell \mathbf{X}_{|g^{(\ell)}}\gamma^{(\ell)}_{|g^{(\ell)}} \\
&\qquad = \sum_\ell \Big[\mathbf{X}_{|g^{(\ell)}}\Big(\hat{\beta}^{(\ell)}_{|g^{(\ell)}} + \frac{1}{\rho}\hat{u}^{(\ell)}_{|g^{(\ell)}}\Big) \\
&\qquad + \frac{1}{\rho}\mathbf{X}_{|g^{(\ell)}}\mathbf{X}^T_{|g^{(\ell)}} y\Big] \qquad (49)\\
&\quad \Rightarrow \quad \sum_\ell \mathbf{X}_{|g^{(\ell)}}\gamma^{(\ell)}_{|g^{(\ell)}} \\
&\qquad = \Big(I + \frac{1}{\rho}\sum_\ell \mathbf{X}_{|g^{(\ell)}}\mathbf{X}^T_{|g^{(\ell)}}\Big)^{-1} \\
&\qquad \cdot \sum_\ell \Big[\mathbf{X}_{|g^{(\ell)}}\Big(\hat{\beta}^{(\ell)}_{|g^{(\ell)}} + \frac{1}{\rho}\hat{u}^{(\ell)}_{|g^{(\ell)}}\Big) \\
&\qquad + \frac{1}{\rho}\mathbf{X}_{|g^{(\ell)}}\mathbf{X}^T_{|g^{(\ell)}} y\Big].
\end{aligned}$$

Substituting (49) into (47) yields

$$\hat{\gamma}^{(\ell)}_{|g^{(\ell)}} = \hat{\beta}^{(\ell)}_{|g^{(\ell)}} + \frac{1}{\rho}\hat{u}^{(\ell)}_{|g^{(\ell)}} + \frac{1}{\rho}\mathbf{X}^T_{|g^{(\ell)}}(y - \Delta),$$

where $\Delta := \sum_\ell \mathbf{X}_{|g^{(\ell)}}\gamma^{(\ell)}_{|g^{(\ell)}}$ in (49).

*Step 2*: Optimize over $\{\beta^{(\ell)}\}$. For $\ell = 1, \dots, L$,

$$\begin{aligned}
\hat{\beta}^{(\ell)} = &\operatorname*{arg\,min}_{\beta^{(\ell)}\in\mathbb{R}^p}\sum_{\ell=1}^{L}\Big\{\frac{1}{2}\Big\|\beta^{(\ell)} - \Big(\hat{\gamma}^{(\ell)} - \frac{1}{\rho}\hat{u}^{(\ell)}\Big)\Big\|_2^2 \\
&+ \frac{\lambda}{\rho}\Omega^{\mathcal{G}_\ell}_{\mathrm{LOG}}(\beta^{(\ell)}; w_{\mathcal{P}_\ell})\Big\} \\
&\text{s.t. } \operatorname{supp}(\beta^{(\ell)}) \subseteq g^{(\ell)},
\end{aligned}$$

$$\hat{\beta}^{(\ell)}_{|g^{(\ell)}} = \mathrm{Prox}^{\mathcal{G}_\ell}_{\mathrm{LOG}}\Big(\Big(\hat{\gamma}^{(\ell)}_{|g^{(\ell)}} - \frac{1}{\rho}\hat{u}^{(\ell)}_{|g^{(\ell)}}\Big); \frac{\lambda}{\rho}, w_{\mathcal{P}_\ell}\Big).$$

All the $\hat{\beta}^{(\ell)}_{|g^{(\ell)}}$'s can be efficiently updated using path-based BCD in Algorithm 4.

*Step 3*: $\hat{u}^{(\ell)} \leftarrow \hat{u}^{(\ell)} + \rho(\hat{\gamma}^{(\ell)} - \hat{\beta}^{(\ell)})$ for $\ell = 1, \dots, L$.

## APPENDIX I: PROOF OF THEOREM 1

If $K = p-1$, then $\hat{K} \le K$.

If $K < p-1$, let $\bar{K}$ be the largest knot such that $\bar{K} \le K$. Then $\hat{K} \ge \bar{K}$. We will show that $\forall k > K$

$$\frac{\|\mathbf{S}_{s_{(\bar{K}+1):k}}\|_F^2}{|s_{(\bar{K}+1):k}|} \le \lambda^2 \qquad (50)$$

through the following two cases.

*Case 1*: If $\bar{K} = K$, then $\forall k > K$, we have

$$\begin{aligned}
\frac{\|\mathbf{S}_{s_{(\bar{K}+1):k}}\|_F^2}{|s_{(\bar{K}+1):k}|} &= \frac{\|\mathbf{S}_{s_{(\bar{K}+1):k}} - \boldsymbol{\Sigma}^*_{s_{(\bar{K}+1):k}}\|_F^2}{|s_{(\bar{K}+1):k}|} \\
&\le \max_{ij}|\mathbf{S}_{ij} - \boldsymbol{\Sigma}^*_{ij}|^2 \le \lambda^2.
\end{aligned} \qquad (51)$$

*Case 2*: If $\bar{K} < K$, then $\forall k > K$, we have

$$\begin{aligned}
\|\mathbf{S}_{s_{(\bar{K}+1):k}}\|_F^2 &= \|\mathbf{S}_{s_{(\bar{K}+1):K}}\|_F^2 \\
&\quad + \|\mathbf{S}_{s_{(K+1):k}} - \boldsymbol{\Sigma}^*_{s_{(K+1):k}}\|_F^2.
\end{aligned} \qquad (52)$$

Since $\bar{K}$ is the largest knot before or at $K$, by Algorithm 7 we have $\forall i = \bar{K}+1, \dots, K$ either (a) or (b) is true:

(a) $\|\mathbf{S}_{s_{(\bar{K}+1):i}}\|_F \le \lambda|s_{(\bar{K}+1):i}|^{1/2}$,

(b) $\exists \bar{k} > i$ s.t. $\|\mathbf{S}_{s_{(\bar{K}+1):i}}\|_F \le \|\mathbf{S}_{s_{(\bar{K}+1):\bar{k}}}\|_F \frac{|s_{(\bar{K}+1):i}|^{1/2}}{|s_{(\bar{K}+1):\bar{k}}|^{1/2}}$.

If (a) holds for $i = K$, then (52) becomes

$$\begin{aligned}
\|\mathbf{S}_{s_{(\bar{K}+1):k}}\|_F^2 &\le \lambda^2|s_{(\bar{K}+1):K}| + \|\mathbf{S}_{s_{(K+1):k}} - \boldsymbol{\Sigma}^*_{s_{(K+1):k}}\|_F^2 \\
&\le \lambda^2|s_{(\bar{K}+1):K}| + \lambda^2|s_{(K+1):k}| \\
&= \lambda^2|s_{(\bar{K}+1):k}|.
\end{aligned}$$

If (b) holds for $i = K$, then $\exists \bar{k} > K$ such that

$$\begin{aligned}
\|\mathbf{S}_{s_{(\bar{K}+1):K}}\|_F^2 &\le \|\mathbf{S}_{s_{(\bar{K}+1):\bar{k}}}\|_F^2\frac{|s_{(\bar{K}+1):K}|}{|s_{(\bar{K}+1):\bar{k}}|} \\
&= (\|\mathbf{S}_{s_{(\bar{K}+1):K}}\|_F^2 + \|\mathbf{S}_{s_{(K+1):\bar{k}}}\|_F^2) \\
&\quad \cdot \frac{|s_{(\bar{K}+1):K}|}{|s_{(\bar{K}+1):\bar{k}}|}.
\end{aligned}$$

Let $\alpha = \frac{|s_{(\bar{K}+1):K}|}{|s_{(\bar{K}+1):\bar{k}}|}$. Then,

$$\begin{aligned}
&\|\mathbf{S}_{s_{(\bar{K}+1):K}}\|_F^2(1-\alpha) \le \|(\mathbf{S} - \boldsymbol{\Sigma}^*)_{(K+1):\bar{k}}\|_F^2\alpha \\
&\quad \Rightarrow \quad \|\mathbf{S}_{s_{(\bar{K}+1):K}}\|_F^2 \le \Big(\frac{\alpha}{1-\alpha}\Big)\lambda^2|s_{(K+1):\bar{k}}|.
\end{aligned} \qquad (53)$$

Let $a = |s_{(\bar{K}+1):K}|$ and $b = |s_{(K+1):\bar{k}}|$. Then $\alpha = \frac{a}{a+b}$. It can be derived that $(\frac{\alpha}{1-\alpha})b = a$. Therefore,

$$\begin{aligned}&\left(\frac{\alpha}{1-\alpha}\right)b \le a\\ &\Rightarrow \quad \left(\frac{\alpha}{1-\alpha}\right)|s_{(K+1):\bar{k}}| \le |s_{(\bar{K}+1):K}|.\end{aligned} \tag{54}$$

Combining (53) and (54) yields

$$\begin{aligned}\|\mathbf{S}_{s_{(\bar{K}+1):K}}\|_F^2 &\le \left(\frac{\alpha}{1-\alpha}\right)\lambda^2|s_{(K+1):\bar{k}}|\\ &\le \lambda^2|s_{(\bar{K}+1):K}|.\end{aligned} \tag{55}$$

Considering $\|\mathbf{S}_{s_{(K+1):k}}\|_F^2 = \|\mathbf{S}_{s_{(K+1):k}} - \boldsymbol{\Sigma}^*_{s_{(K+1):k}}\|_F^2 \le \lambda^2|s_{(K+1):k}|$ and (55), we have

$$\|\mathbf{S}_{s_{(\bar{K}+1):k}}\|_F^2 \le \lambda^2|s_{(\bar{K}+1):k}|.$$

In both Case 1 and Case 2, we have $\frac{\|\mathbf{S}_{s_{(\bar{K}+1):k}}\|_F^2}{|s_{(\bar{K}+1):k}|} \le \lambda^2$. By Algorithm 7, $\bar{K}$ is the last knot in both cases. Hence, $\hat{K} = \bar{K} \le K$.

## APPENDIX J: PROOF OF THEOREM 2

Let $\tilde{K}$ be the largest knot such that $\tilde{K} < K$. Being on the set $\mathcal{A}_x$ implies that, for any $k > \tilde{K}$,

$$\begin{aligned}\|\mathbf{S}_{s_{(\tilde{K}+1):k}}\|_F &\ge \|\boldsymbol{\Sigma}^*_{s_{(\tilde{K}+1):k}}\|_F\\ &\quad - \|\mathbf{S}_{s_{(\tilde{K}+1):k}} - \boldsymbol{\Sigma}^*_{s_{(\tilde{K}+1):k}}\|_F\\ &\ge \|\boldsymbol{\Sigma}^*_{s_{(\tilde{K}+1):k}}\|_F - \lambda\sqrt{|s_{(\tilde{K}+1):k}|}.\end{aligned} \tag{56}$$

From (56), we have

$$\begin{aligned}\max_{k\ge K}\left\{\frac{\|\mathbf{S}_{s_{(\tilde{K}+1):k}}\|_F}{|s_{(\tilde{K}+1):k}|^{\frac{1}{2}}}\right\} &\ge \max_{k\ge K}\left\{\frac{\|\boldsymbol{\Sigma}^*_{s_{(\tilde{K}+1):k}}\|_F}{|s_{(\tilde{K}+1):k}|^{\frac{1}{2}}}\right\} - \lambda\\ &\ge \frac{\|\boldsymbol{\Sigma}^*_{s_{(\tilde{K}+1):K}}\|_F}{|s_{(\tilde{K}+1):K}|^{\frac{1}{2}}} - \lambda\\ &> 2\lambda - \lambda = \lambda.\end{aligned} \tag{57}$$

where the last equality holds by Assumption (21), given $\tilde{K} + 1 \le K$. Equivalently, $\exists k \ge K$ such that

$$\frac{\|\mathbf{S}_{s_{(\tilde{K}+1):k}}\|_F^2}{|s_{(\tilde{K}+1):k}|} > \lambda^2. \tag{58}$$

There exists a knot $k \ge K$ when applying Algorithm 7 to solve the problem. Hence, $\hat{K} \ge K$.

## APPENDIX K: PROOF OF THEOREM 3

We can rewrite Problem (20) in terms of the latent variables $\{\mathbf{V}^{(k)}\}_{k=1}^{p-1}$:

$$\begin{aligned}\{\hat{\mathbf{V}}^{(k)}\}_{k=1}^{p-1} = \operatorname*{arg\,min}_{\mathbf{V}^{(1)},\dots,\mathbf{V}^{(p-1)}\in\mathbb{R}^{p\times p}} &\left\{\frac{1}{2}\left\|\sum_{k=1}^{p-1}\mathbf{V}^{(k)} - \mathbf{S}^-\right\|_F^2\right.\\ &+ \lambda\sum_{k=1}^{p-1} w_k\|\mathbf{V}^{(k)}\|_F\\ &\left.\text{s.t. } \operatorname{supp}(\mathbf{V}^{(k)}) \subseteq s_{1:k}\right\}\end{aligned} \tag{59}$$

so that $\hat{\boldsymbol{\Sigma}}^{\text{LOG}-} = \sum_{k=1}^{p-1}\hat{\mathbf{V}}^{(k)}$. In addition, $\hat{\boldsymbol{\Sigma}}^{\text{LOG}}_{s_0} = \mathbf{S}_{s_0}$ because the LOG penalty does not apply to the diagonal elements. Taking subgradient of the objective function in (59) with respect to $\mathbf{V}^{(K)}$ where $K$ is the bandwidth of $\boldsymbol{\Sigma}^*$ yields

$$0 \in \left(\sum_{k=1}^{p-1}\hat{\mathbf{V}}^{(k)} - \mathbf{S}^-\right)_{s_{1:K}} + \lambda w_K\partial\|\mathbf{V}^{(K)}\|_F. \tag{60}$$

When $\mathbf{V}^{(K)} \ne 0$,

$$\partial\|\mathbf{V}^{(K)}\|_F = \frac{\mathbf{V}^{(K)}}{\|\mathbf{V}^{(K)}\|_F}. \tag{61}$$

When $\mathbf{V}^{(K)} = 0$,

$$\begin{aligned}\partial\|\mathbf{V}^{(K)}\|_F &= \{\mathbf{Z}\in\mathbb{R}^{p\times p}:\\ &\qquad \|\mathbf{U}\|_F \ge \|\mathbf{V}^{(K)}\|_F + \langle\mathbf{Z}, \mathbf{U} - \mathbf{V}^{(k)}\rangle\\ &\qquad \forall\mathbf{U}\in\mathbb{R}^{p\times p}\}\\ &= \{\mathbf{Z}\in\mathbb{R}^{p\times p}: \|\mathbf{U}\|_F \ge \langle\mathbf{Z}, \mathbf{U}\rangle\\ &\qquad \forall\mathbf{U}\in\mathbb{R}^{p\times p}\}\\ &= \{\mathbf{Z}\in\mathbb{R}^{p\times p}: \|\mathbf{Z}\|_F \le 1\}.\end{aligned} \tag{62}$$

Combining (60), (61) and (62) we have

$$\begin{aligned}&\left\|\left(\sum_{k=1}^{p-1}\hat{\mathbf{V}}^{(k)} - \mathbf{S}^-\right)_{s_{1:K}}\right\|_F \le \lambda w_K\\ &\Leftrightarrow \quad \|(\hat{\boldsymbol{\Sigma}}^{\text{LOG}-} - \mathbf{S}^-)_{s_{1:K}}\|_F \le \lambda w_K\\ &\Leftrightarrow \quad \|(\hat{\boldsymbol{\Sigma}}^{\text{LOG}} - \mathbf{S})_{s_{1:K}}\|_F \le \lambda w_K.\end{aligned} \tag{63}$$

Furthermore, on $\mathcal{A}_x$ we have

$$\lambda^2 \ge \max_{i=j}|\mathbf{S}_{ij} - \boldsymbol{\Sigma}^*_{ij}|^2 \ge \frac{1}{p}\|\mathbf{S}_{s_0} - \boldsymbol{\Sigma}^*_{s_0}\|_F^2, \tag{64}$$

$$\lambda \ge \max_{i,j}|\mathbf{S}_{ij} - \boldsymbol{\Sigma}^*_{ij}| \ge \frac{1}{\sqrt{|s_{1:K}|}}\|(\mathbf{S} - \boldsymbol{\Sigma}^*)_{s_{1:K}}\|_F. \tag{65}$$

Using triangle inequality, (63) and (65) we have

$$\begin{aligned}
&\|(\hat{\mathbf{\Sigma}}^{\mathrm{LOG}} - \mathbf{\Sigma}^*)_{s_{1:K}}\|_F \\
&\le \|(\hat{\mathbf{\Sigma}}^{\mathrm{LOG}} - \mathbf{S})_{s_{1:K}}\|_F + \|(\mathbf{S} - \mathbf{\Sigma}^*)_{s_{1:K}}\|_F \\
&\le \lambda w_K + \lambda\sqrt{|s_{1:K}|} \\
&= 2\lambda\sqrt{|s_{1:K}|}.
\end{aligned} \tag{66}$$

Using (64) and (66) we have

$$\begin{aligned}
\|\hat{\mathbf{\Sigma}}^{\mathrm{LOG}} - \mathbf{\Sigma}^*\|_F^2 &= \|(\hat{\mathbf{\Sigma}}^{\mathrm{LOG}} - \mathbf{\Sigma}^*)_{s_{1:K}}\|_F^2 \\
&\quad + \|\hat{\mathbf{\Sigma}}^{\mathrm{LOG}}_{s_0} - \mathbf{\Sigma}^*_{s_0}\|_F^2 \\
&= \|(\hat{\mathbf{\Sigma}}^{\mathrm{LOG}} - \mathbf{\Sigma}^*)_{s_{1:K}}\|_F^2 \\
&\quad + \|\mathbf{S}_{s_0} - \mathbf{\Sigma}^*_{s_0}\|_F^2 \\
&\le 4\lambda^2|s_{1:K}| + \lambda^2 p \\
&\le \frac{4x^2 pK\log p}{n} + \frac{x^2 p\log p}{n}.
\end{aligned} \tag{67}$$

By Theorem 1, $\hat{K} \le K$ with high probability when $\lambda \ge x\sqrt{\log p/n}$. Therefore, the equality in (67) holds with high probability. Hence,

$$\|\hat{\mathbf{\Sigma}}^{\mathrm{LOG}} - \mathbf{\Sigma}^*\|_F^2 \lesssim pK\log p/n.$$

## APPENDIX L: ALGORITHM 7 FOR SOLVING (20), MODIFIED FROM ALGORITHM 3

**Algorithm 7** Solve for $\hat{\mathbf{\Sigma}}^{\mathrm{LOG}}$ defined by Problem (20)

**Input:** $\lambda \ge 0$, $\mathbf{S} \in \mathbb{R}^{p\times p}$ and $a(\mathcal{D})$.

```
Σ ← S_{s_0}
k ← 0
while k < p − 1 do
    K ← arg max_{j:j>k} f(j, k)
      ▷ f(j, k) = ‖S_{s_{(k+1):j}}‖_F / sqrt(|s_{(k+1):j}|)  for 0 ≤ k < j ≤ p − 1
    if f(K, k) ≤ λ then
        break
    end if
    Σ_{s_{(k+1):K}} ← S_G(S_{s_{(k+1):K}}, λ sqrt(|s_{(k+1):K}|))
    k ← K
end while
```

**Output:** $\mathbf{\Sigma}$

## APPENDIX M: PSD PROBABILITY (FIGURE 12) AND MINIMUM EIGENVALUES (FIGURE 13) OF THE THREE COVARIANCE ESTIMATORS

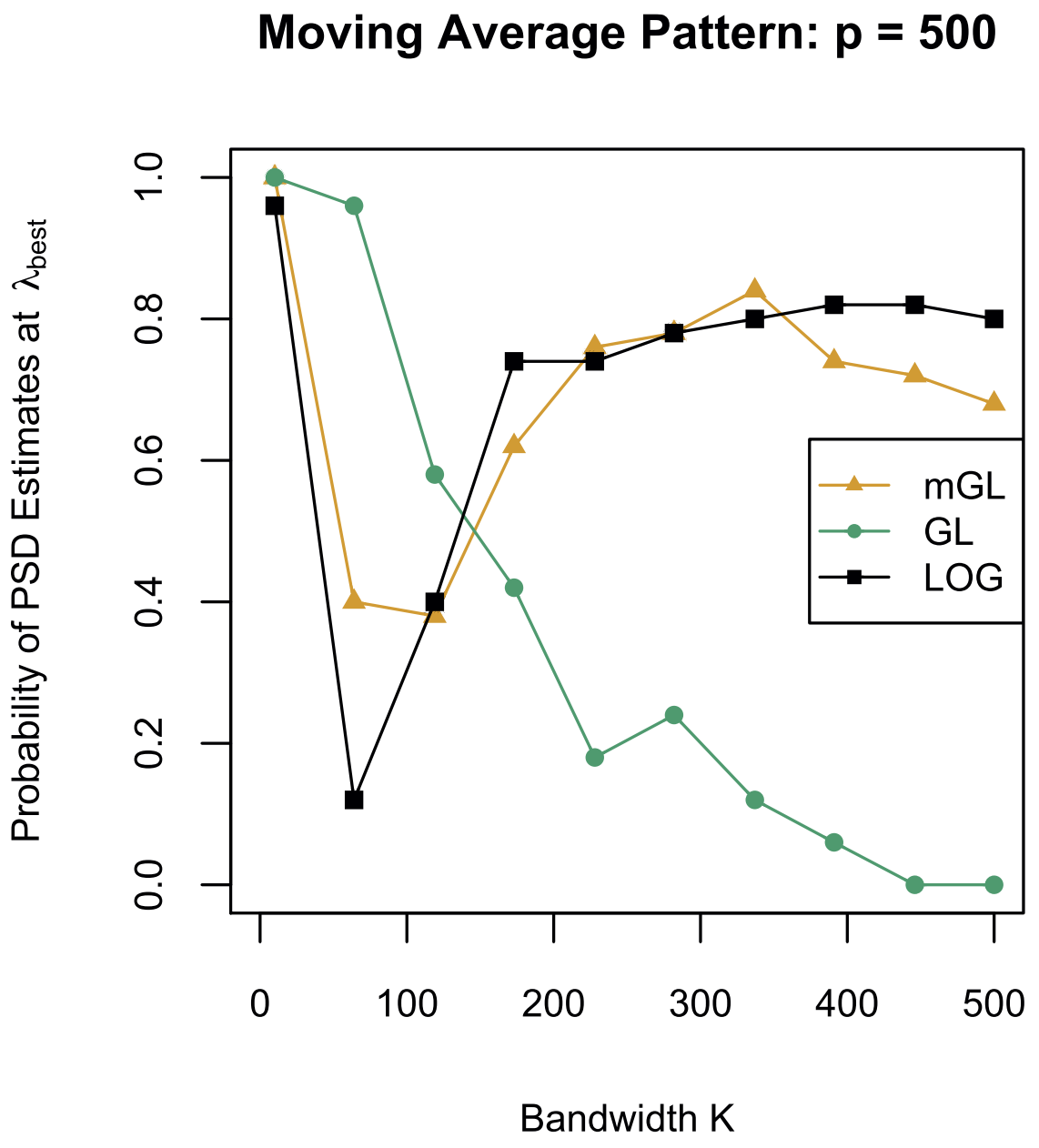

FIG. 12. *For the three estimators* ($\hat{\mathbf{\Sigma}}^{\mathrm{mGL}}, \hat{\mathbf{\Sigma}}^{\mathrm{GL}}, \hat{\mathbf{\Sigma}}^{\mathrm{LOG}}$) *in moving-average pattern, probability of their estimates being PSD at* $\lambda_{\mathrm{best}}$.

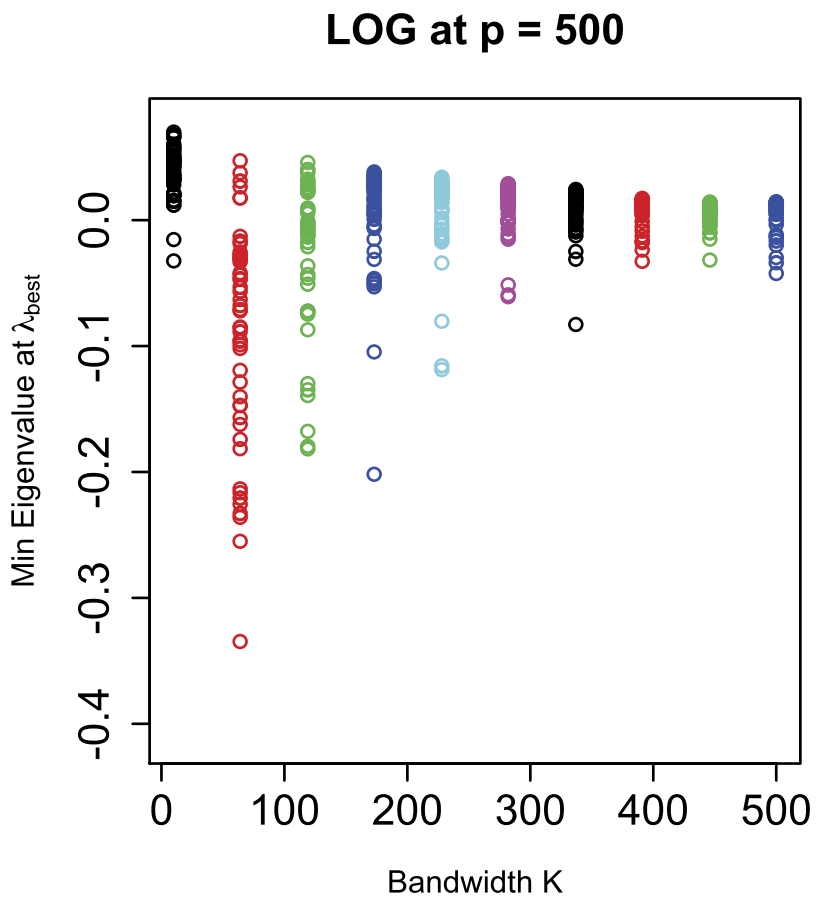

FIG. 13. *For the three estimators* ($\hat{\mathbf{\Sigma}}^{\mathrm{LOG}}, \hat{\mathbf{\Sigma}}^{\mathrm{mGL}}, \hat{\mathbf{\Sigma}}^{\mathrm{GL}}$) *in moving-average pattern, minimum eigenvalues of* 50 *samples at* $\lambda_{\mathrm{best}}$.

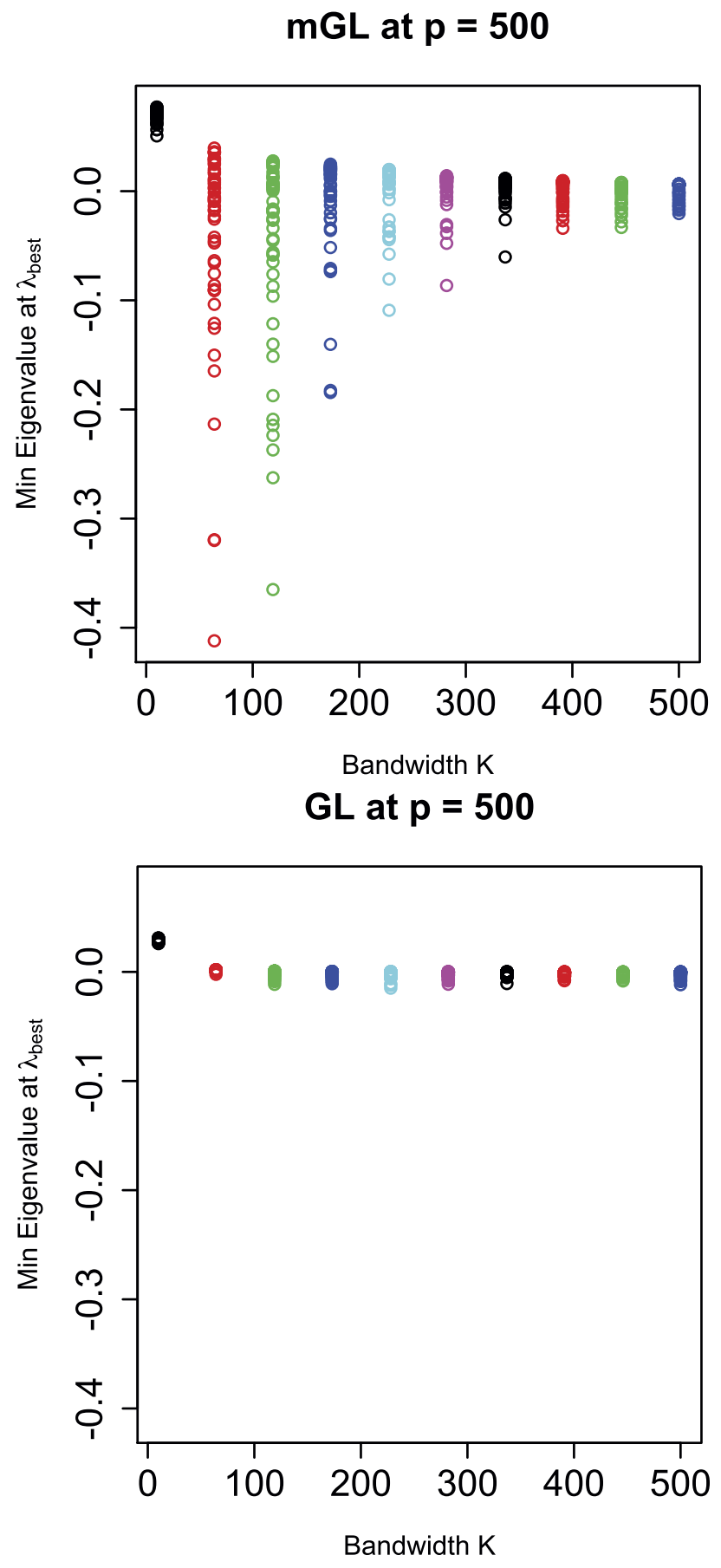

FIG. 13. (*Continued*).

## ACKNOWLEDGMENT

This work was supported by NSF DMS-1405746.

## REFERENCES

BACH, F. (2008). Exploring large feature spaces with hierarchical multiple kernel learning. In *Proceedings of the* 21*st International Conference on Neural Information Processing Systems*. *NIPS'*08 105–112. Curran Associates Inc., Red Hook, NY.

BACH, F., JENATTON, R., MAIRAL, J. and OBOZINSKI, G. (2012). Structured sparsity through convex optimization. *Statist. Sci*. **27** 450–468. MR3025128

BECK, A. and TEBOULLE, M. (2009). A fast iterative shrinkage-thresholding algorithm for linear inverse problems. *SIAM J. Imaging Sci*. **2** 183–202. MR2486527

BERTSEKAS, D. P. (1999). *Nonlinear Programming*, 2nd ed. Athena Scientific, Belmont, MA. MR3444832

BIEN, J., BUNEA, F. and XIAO, L. (2016). Convex banding of the covariance matrix. *J. Amer. Statist. Assoc*. **111** 834–845. MR3538709

BIEN, J., TAYLOR, J. and TIBSHIRANI, R. (2013). A LASSO for hierarchical interactions. *Ann. Statist*. **41** 1111–1141. MR3113805

BOYD, S. and VANDENBERGHE, L. (2004). *Convex Optimization*. Cambridge Univ. Press, Cambridge. MR2061575

BOYD, S., PARIKH, N., CHU, E., PELEATO, B. and ECKSTEIN, J. (2011). Distributed optimization and statistical learning via the alternating direction method of multipliers. *Found. Trends Mach. Learn*. **3** 1–122.

CHOI, N. H., LI, W. and ZHU, J. (2010). Variable selection with the strong heredity constraint and its oracle property. *J. Amer. Statist. Assoc*. **105** 354–364. MR2656056

CHOULDECHOVA, A. and HASTIE, T. (2015). Generalized additive model selection. Available at arXiv:1506.03850.

HARIS, A., WITTEN, D. and SIMON, N. (2016). Convex modeling of interactions with strong heredity. *J. Comput. Graph. Statist*. **25** 981–1004. MR3572025

JACOB, L., OBOZINSKI, G. and VERT, J. (2009). Group lasso with overlap and graph lasso. In *Proceedings of the* 26*th Annual International Conference on Machine Learning*. *ICML'*09 433–440. ACM, New York.

JENATTON, R., AUDIBERT, J.-Y. and BACH, F. (2011). Structured variable selection with sparsity-inducing norms. *J. Mach. Learn. Res*. **12** 2777–2824. MR2854347

JENATTON, R., MAIRAL, J., OBOZINSKI, G. and BACH, F. (2010). Proximal methods for sparse hierarchical dictionary learning. In *Proceedings of the* 27*th International Conference on International Conference on Machine Learning*. *ICML'*10 487–494. Omnipress, Madison, WI.

JENATTON, R., MAIRAL, J., OBOZINSKI, G. and BACH, F. (2011). Proximal methods for hierarchical sparse coding. *J. Mach. Learn. Res*. **12** 2297–2334. MR2825428

LEVINA, E., ROTHMAN, A. and ZHU, J. (2008). Sparse estimation of large covariance matrices via a nested Lasso penalty. *Ann. Appl. Stat*. **2** 245–263. MR2415602

LIM, M. and HASTIE, T. (2015). Learning interactions via hierarchical group-lasso regularization. *J. Comput. Graph. Statist*. **24** 627–654. MR3397226

LOU, Y., BIEN, J., CARUANA, R. and GEHRKE, J. (2016). Sparse partially linear additive models. *J. Comput. Graph. Statist*. **25** 1026–1040. MR3572032

NELDER, J. A. (1977). A reformulation of linear models. *J. Roy. Statist. Soc. Ser. A* **140** 48–76. MR0458743

NESTEROV, YU. (2013). Gradient methods for minimizing composite functions. *Math. Program*. **140** 125–161. MR3071865

NICHOLSON, W. B., BIEN, J. and MATTESON, D. S. (2014). Hierarchical vector autoregression. Available at arXiv:1412.5250.

OBOZINSKI, G., JACOB, L. and VERT, J.-P. (2011). Group lasso with overlaps: The latent group lasso approach. Research report. Available at https://hal.inria.fr/inria-00628498.

RADCHENKO, P. and JAMES, G. M. (2010). Variable selection using adaptive nonlinear interaction structures in high dimensions. *J. Amer. Statist. Assoc*. **105** 1541–1553. MR2796570

ROTHMAN, A. J., LEVINA, E. and ZHU, J. (2010). A new approach to Cholesky-based covariance regularization in high dimensions. *Biometrika* **97** 539–550. MR2672482

SCHMIDT, M. and MURPHY, K. (2010). Convex structure learning in log-linear models: Beyond pairwise potentials. In *Proceedings of the* 13*th International Conference on Artificial Intelligence and Statistics*. *Proceedings of Machine Learning Research* **9** 709–716.

SHE, Y., WANG, Z. and JIANG, H. (2017). Group regularized estimation under structural hierarchy. *J. Amer. Statist. Assoc*. To appear. DOI:10.1080/01621459.2016.1260470.

SIMON, N., FRIEDMAN, J., HASTIE, T. and TIBSHIRANI, R. (2013). A sparse-group lasso. *J. Comput. Graph. Statist.* **22** 231–245. MR3173712

TIBSHIRANI, R. (1996). Regression shrinkage and selection via the lasso. *J. Roy. Statist. Soc. Ser. B* **58** 267–288. MR1379242

TSENG, P. (2001). Convergence of a block coordinate descent method for nondifferentiable minimization. *J. Optim. Theory Appl.* **109** 475–494. MR1835069

TURLACH, B. A., VENABLES, W. N. and WRIGHT, S. J. (2005). Simultaneous variable selection. *Technometrics* **47** 349–363. MR2164706

VILLA, S., ROSASCO, L., MOSCI, S. and VERRI, A. (2014). Proximal methods for the latent group lasso penalty. *Comput. Optim. Appl.* **58** 381–407. MR3201966

YUAN, M., JOSEPH, V. R. and ZOU, H. (2009). Structured variable selection and estimation. *Ann. Appl. Stat.* **3** 1738–1757. MR2752156

YUAN, M. and LIN, Y. (2006). Model selection and estimation in regression with grouped variables. *J. Roy. Statist. Soc. Ser. B* **68** 49–67. MR2212574

ZHAO, P., ROCHA, G. and YU, B. (2009). The composite absolute penalties family for grouped and hierarchical variable selection. *Ann. Statist.* **37** 3468–3497. MR2549566